\documentstyle[12pt,twoside]{article}
\newcommand{\als}{$\alpha_s$}

\def\lsim{\mathrel{\rlap{\lower4pt\hbox{\hskip1pt$\sim$}}
    \raise1pt\hbox{$<$}}}         
\def\gsim{\mathrel{\rlap{\lower4pt\hbox{\hskip1pt$\sim$}}
    \raise1pt\hbox{$>$}}}         

\newlength{\dinwidth}
\newlength{\dinmargin}
  \newenvironment{defl}[1]%
  {\begin{list}{}{\settowidth{\labelwidth}{#1}%
  \setlength{\leftmargin}{\labelwidth}%
  \addtolength{\leftmargin}{\labelsep}%
  \setlength{\itemsep}{0pt plus 1pt}
  \setlength{\parsep}{0pt plus 1pt}
  \setlength{\topsep}{0pt plus 1pt}
  \setlength{\partopsep}{0pt plus 1pt}
  \setlength{\parskip}{2mm plus 1mm minus 1mm}
  }}%
  {\end{list}}
\begin{document}
\thispagestyle{empty}   
\noindent
DESY 96--057          \hfill ISSN 0418-9833\\
April 1996        
\begin{center}
  \begin{Large}
  \begin{bf}
LEPTO\footnote{Information and code via WWW page 
\texttt{http://www3.tsl.uu.se/thep/lepto/}}
6.5 --- A Monte Carlo Generator\\
for Deep Inelastic Lepton-Nucleon Scattering\\
  \end{bf}
  \end{Large}
  \vspace{5mm}
  \begin{large}
    G. Ingelman$^{a,b}$, A. Edin$^b$, J. Rathsman$^b$\\
  \end{large}
  \vspace{3mm}
ingelman@desy.de ~~~~ rathsman@tsl.uu.se ~~~~ edin@tsl.uu.se\\
  \vspace{3mm}
$^a$ Deutsches~Elektronen-Synchrotron~DESY,
Notkestrasse~85,~D-22603~Hamburg,~FRG\\
$^b$ Dept. of Radiation Sciences, Uppsala University,
Box 535, S-751 21 Uppsala, Sweden\\
\vspace{5mm}
\end{center}
\begin{quotation}
\noindent

{\bf Abstract.}
Physics and programming aspects are discussed for a Fortran 77 Monte Carlo 
program to simulate complete events in deep inelastic lepton-nucleon
scattering. The parton level interaction is based on the standard model
electroweak cross sections, which are fully implemented in leading order for
any lepton of arbitrary polarization, and different parametrizations of parton
density functions can be used. First order QCD matrix elements for gluon
radiation and boson-gluon fusion are implemented and higher order QCD radiation
is treated using parton showers. Hadronization is performed using the Lund
string model, implemented in {\sc Jetset}/{\sc Pythia}. Rapidity gap events are generated
through a model based on soft  colour interactions.
\end{quotation}

\section{Introduction}
\label{sec:intro}
Deep inelastic lepton-nucleon scattering \cite{EXPREV} has played an important
role for probing the structure of the proton and understanding the fundamental
electro\-magnetic, weak and strong interactions at the level of quarks and
leptons. With the order of magnitude increase in the centre-of-mass energy now
available through $ep$ collisions in HERA, this line of research will continue
to be at the forefront.

The basic lepton-quark scattering processes have well-defined cross section
formulae within the electroweak standard model \cite{EWSM}. With the inclusion
of parton density distributions and perturbative QCD corrections the problem of
practical evaluations become quite complex and analytical calculations are only
possible in simplified cases or through approximations. Normal numerical
methods are in many cases possible, but Monte Carlo simulation is often
preferable because of its generality and applicability to complex problems. In
the case of the multiparticle hadronic final state, the only viable alternative
is, in fact, the Monte Carlo method.

The present program, {\sc Lepto}, is a general and flexible Monte Carlo (MC) to
simulate complete lepton-nucleon scattering events and integrate cross
sections. It is based on the leading order electro\-weak cross sections for the
underlying parton level scattering processes, and includes QCD corrections
using exact first order matrix elements and higher orders in the leading $\log
Q^2$ parton cascade approach. The fragmentation of produced partons into
observable hadrons is performed with the Lund string hadronization model
\cite{LUND}. An arbitrary configuration of a lepton and a nucleon can be
defined with constraints on the scattering kinematics and the generated events
can be transformed to different frames. The present version is the latest
development in a series of earlier versions \cite{LEPTO,LEPTO61} that have
developed in stages and profitted greatly by feed-back from extensive
comparisons with experimental data on the hadronic final state in
leptoproduction. This has given valuable insights concerning the QCD processes
of gluon radiation and boson-gluon fusion as well as the confinement induced
hadronization of colour charged partons. The generally good agreement between
data, in particular from the European Muon Collaboration and the neutrino
experiments WA21 and WA25 at CERN, and the Monte Carlo shows the validity of
the models and procedures used in the program. This can be illustrated by the
results on longitudinal momentum spectra of different identified hadrons
\cite{XFDISTR}, transverse momentum properties, energy flows and jet structure
\cite{PTEFF}. The much higher  energies available in $ep$ collisions at HERA
provides both qualitative  and quantitative new information on QCD effects,
such that detailed tests of MC models are now being performed
\cite{H1-LEPTO,ZEUS-LEPTO}.

In the following a comprehensive description is given of the theoretical
framework built into the program (section 2) as well as the various program
components (section 3) and their usage (section 4).


\section{Physics and MC implementation}
\subsection{Kinematics}
The main kinematic relations given here are for the case of electron-proton
scattering, $e+p\rightarrow \ell + H$ where $\ell$
is the scattered lepton and $H$ the final hadron system,
but are of course equally valid for any lepton beam or a neutron
target. Let $p_e,\; p_{\ell}$ be the
four-vectors of the incoming and scattered lepton, respectively,
and $P$ that of the
incoming proton. Some basic kinematic relations are then
(cf.\ \cite{DIPSREP,EPKIN})
\begin{eqnarray} \label{KINREL}
  s & \equiv & (p_e + P)^2 \simeq 4E_e E_p \\
  W^2 & \equiv & (q + P)^2 = Q^2 \frac{1-x}{x} + m_p^2 \\
  Q^2 & \equiv & -q^2 = -(p_e - p_{\ell})^2
        \simeq 4 E_e E_{\ell}\sin^2{\frac{\theta_{\ell}}{2}} \\
  m_p\nu & \equiv & P\cdot q
     \simeq 2 E_p (E_e - E_{\ell}\cos^2{\frac{\theta_{\ell}}{2}}) \\
  x & \equiv & \frac{Q^2}{2P\cdot q} = \frac{Q^2}{2m_p\nu } \simeq
      \frac{E_e E_{\ell} \sin^2{\frac{\theta_{\ell}}{2}}}{E_p
               (E_e - E_{\ell}\cos^2{\frac{\theta_{\ell}}{2}})} \\
  y & \equiv & \frac{P\cdot q}{P\cdot p_e}
      = \frac{\nu}{\nu_{max}}
      \simeq \frac{E_e - E_{\ell}\cos^2{\frac{\theta_{\ell}}{2}}}{E_e}
\end{eqnarray}
where the `$\simeq$' sign means that the lepton and nucleon masses are
neglected, an approximation which is usually acceptable already at
fixed target energies. Nevertheless, exact formulae are used in the 
program to avoid unnecessary approximations. 
Here, $\sqrt{s}$ is the total invariant mass and $W$ the invariant mass
of the hadronic system $H$.
The exchanged vector boson, $\gamma/Z^0$ for neutral
current interactions and $W^{\pm}$ for charged current interactions,
carries the momentum transfer $Q^2$ and the variable $\nu$
is the energy of this current in the target rest frame.
Finally, Bjorken-$x$ and $y$ are the convenient dimensionless scaling variables
in the range $[0,1]$.

For the overall event kinematics there are only two
independent variables and hence by measuring, e.g., the
energy ($E_{\ell}$) and angle ($\theta_{\ell}$) of the scattered lepton 
(here defined with respect to the incoming electron direction), 
all other global variables can be calculated.
For theoretical purposes, cross section formulae etc, the variables
($x,y$) or ($x,Q^2$) are usually used as the independent ones.
No assumption has here been made about the structure
of the proton, or about the final hadronic state.
By {\it assuming} the quark-parton model, where the
current couples to a quark with four-vector
$p_i \equiv \xi (E_p,0,0,E_p)$ and {\it assuming} the initial and
final quark to be massless one finds $\xi = x$.
The Bjorken-$x$ variable can then be interpreted as the momentum
fraction of the proton which is carried by the struck quark. However,
taking QCD corrections into account this relation no longer
holds \cite{PS}.

For $ep$ colliders like HERA where the proton beam energy is much larger
than the electron beam energy, the phase space is somewhat special and quite
elongated in the proton beam direction (see Fig.~2 in \cite{DIPSREP}).
Although the events are certainly not evenly
distributed in the available phase space, they are
in general very asymmetric, with most of the final state
hadrons in the `forward' direction along the incoming proton.

For most analyses the event kinematics need to be reconstructed. The
differential cross sections, which are the basis of most analyses, are usually
needed as functions of, e.g., $x$ and $Q^2$. The event kinematics can be
straightforwardly obtained from the formulae above if the scattered lepton can
be well measured. However, this is not always the case due to instrumental
effects or because it is a neutrino. These problems are accentuated at $ep$
colliders where the acceptance is limited by the beam pipes and where charged
current interactions give an undetectable neutrino as scattered lepton. This
implies the importance of being able to use the hadronic part of the event to
reconstruct the  kinematical variables. In the naive quark-parton model (QPM)
the scattered quark gives the necessary information, but in reality a number of
smearing effects enter (QCD processes, fragmentation, mass effects, jet
reconstruction). Considering the hadronic final state as a single system, whose
internal structure is of no importance, and applying energy momentum
conservation between this system and the scattered lepton leads to  methods
such as `Jacquet--Blondel' \cite{BIS,JB}, `double angle' \cite{DA}  and
`$\Sigma$' \cite{SIGMA}. These are particularly suited to $ep$ colliders, since
they minimize the effects of particles lost in the beam pipe.

\subsection{Electroweak cross sections}
In leading order electroweak theory \cite{EWSM}
the differential neutral current (NC) cross section
for the scattering of a charged lepton is (neglecting masses) given by
\begin{equation}
\label{NC}
 \frac{d^2\sigma_{NC}(e^{\mp})}{dx\: dQ^2} = \frac{4\pi \alpha^2}{xQ^4}
\left[ y^2 xF_1(x,Q^2) + (1-y) F_2(x,Q^2)
\pm \left( y-\frac{y^2}{2} \right) xF_3(x,Q^2) \right]
\end{equation}
in terms of the nucleon structure functions $F_1,F_2,F_3$.
Two of these are related through
the Callan-Gross relation, $2xF_1=F_2$, which holds for spin 1/2 quarks
when neglecting masses, intrinsic transverse momenta and order
\als \ QCD effects.
These effects are usually negligibly small, except at small-$x$ 
\cite{KLEIN,MANDY}, but can optionally be included (cf.\ section 2.3). 
More important for the structure of the electroweak
theory is to take the lepton beam polarization into account.
For a left- ($L$) and right-handed ($R$) electron one has the
differential cross section
\begin{equation} \label{NCPOL}
 \frac{d^2\sigma_{NC}(e^-_{L,R}p)}{dx\: dQ^2} = \frac{2\pi \alpha^2}{xQ^4}
\left[ \left(1+(1-y)^2\right) F_2^{L,R}(x,Q^2)
      +\left(1-(1-y)^2\right) xF_3^{L,R}(x,Q^2)\right]
\end{equation}
The structure functions are in the standard parton model given by
\begin{eqnarray}
  \label{F2F3}
F_2^{L,R}(x,q^2) & = &\sum_f \left[ xq_f(x,Q^2)+x\bar{q}_f(x,Q^2)\right]
                     \;  A_f^{L,R}(Q^2) \\
xF_3^{L,R}(x,q^2)& = &\sum_f \left[ xq_f(x,Q^2)-x\bar{q}_f(x,Q^2)\right]
                     \;  B_f^{L,R}(Q^2)
\end{eqnarray}
where the sum is over all flavours $f$ and
$q_f(\bar{q}_f)$ denote the quark (antiquark) density distributions
in the nucleon (cf.\ section 2.4).
The coefficients are given by
\begin{eqnarray}
  \label{ACOEFF}
A_f^{L,R}(Q^2) & = & e_f^2 - 2e_f(v_e\pm a_e)v_f P_Z +
                               (v_e\pm a_e)^2 (v_f^2+a_f^2) P_Z^2 \\
  \label{BCOEFF}
B_f^{L,R}(Q^2) & = & \;\;\;\; \mp \; 2e_f(v_e\pm a_e)a_f P_Z
                            \pm 2(v_e\pm a_e)^2 v_fa_f P_Z^2
\end{eqnarray}
where $e_f$ is the electric charge $(e_e=-1)$,
$v_f = (T_{3f}-2e_f \sin^2{\theta_W})/\sin{2\theta_W}$ and
$a_f = T_{3f}/\sin{2\theta_W}$
are the NC vector and axial vector couplings
expressed in terms of the third component of the weak isospin
$(T_{3e}=-\frac{1}{2})$ and the Weinberg angle $\theta_{W}$.
$P_{Z}$ is the ratio of the $Z$ and $\gamma$ propagators
$P_Z = Q^2/(Q^2+M_Z^2)$.
The corresponding cross sections for left- and right-handed positrons,
$e^+_{L,R}$, are obtained from the above electron formulae
by the replacements
\begin{equation}
  \label{REP}
 F_2^{L,R} \rightarrow F_2^{R,L}, \; \;
 xF_3^{L,R} \rightarrow -xF_3^{R,L}
\end{equation}
The cross section for an arbitrarily polarized electron/positron beam
is simply obtained as a linear combination of these pure left- and
right-handed cross sections (cf.\ \cite{KLEIN}).

The pure $\gamma$ exchange term, i.e.\ the one without a $P_Z$
dependence in eq. (\ref{ACOEFF}), dominates completely at low $Q^2$,
and the
cross section then takes the familiar form measured in
fixed target electron and muon beam experiments
\begin{equation} \label{SIGMAEM}
 \frac{d^2\sigma_{\gamma}(ep)}{dx\: dQ^2} =
 \frac{2\pi \alpha^2}{xQ^4}
 (1+(1-y)^2) F_2^{em}(x,Q^2)
\end{equation}
where the electromagnetic structure function is given by
\begin{equation}   \label{F2EM}
 F_2^{em}(x,Q^2) = \sum_f e_f^2 \left[ xq_f(x,Q^2) + x\bar{q}_f(x,Q^2) \right]
\end{equation}
With increasing $Q^2$ first the $\gamma /Z^0$ interference
term (linear in $P_Z$) and then the pure weak term (quadratic in $P_Z$)
become important and finally dominate the cross section at large
$Q^2$, see e.g. Fig.~3 in \cite{DIPSREP}.

The differential cross sections for charged current (CC) $ep$
interactions are
given by
\begin{eqnarray}
\label{CCMINUS}
\frac{d^2\sigma_{CC}(e^-p)}{dx\: dQ^2} & = &
\frac{(1-\lambda )\pi \alpha^2}{4\sin^4{\theta_W}
                                     \left( Q^2+M_W^2\right)^2}
\sum_{i,j} \left[ \left| V_{u_id_j} \right|^2 u_i +
      (1-y)^2 \left| V_{u_jd_i} \right|^2 \bar{d}_i \right] \\
\label{CCPLUS}
\frac{d^2\sigma_{CC}(e^+p)}{dx\: dQ^2} & = &
\frac{(1+\lambda )\pi \alpha^2}{4\sin^4{\theta_W}
                                     \left( Q^2+M_W^2\right)^2}
\sum_{i,j} \left[ \left| V_{u_id_j} \right|^2 \bar{u}_i +
      (1-y)^2 \left| V_{u_jd_i} \right|^2 d_i \right]
\end{eqnarray}
where $V_{u_id_j}$ are elements of the Kobayashi-Maskawa
matrix, $u_i$ and $d_j$ denote the parton density functions for the
up-type and down-type quark flavours,
respectively, and $i,j$ are family indices.
The $e^{\pm}$ beam polarization is denoted by $\lambda$
($\pm 1$ for a right/left-handed state).
Considering only four massless quark flavours ($u,d,s,c$) and using
the unitarity relation
$\sum_j \left| V_{u_id_j}\right|^2=\sum_j \left| V_{u_jd_i}\right|^2=1$
one obtains for any lepton with fixed helicity
\begin{equation}
  \label{CCGF}
 \frac{d^2\sigma_{CC}(\ell p)}{dx \: dQ^2 } \simeq \frac{G_F^2}{\pi}
           \left( 1+\frac{Q^2}{M_W^2} \right)^{-2} \left\{
  \begin{array}{ll}
  (u+c) + (1-y)^2(\bar{d}+\bar{s}) & for \;\;\;\; \ell = e_L^-, \; \bar{\nu} \\
  (\bar{u}+\bar{c}) + (1-y)^2(d+s) & for \;\;\;\; \ell = e_R^+, \; \nu       \\
  0                                & for \;\;\;\; \ell = e_R^-, \; e_L^+
  \end{array}
  \right.
\end{equation}
Here, $G_F=\pi\alpha/(\sqrt{2}\sin^2{\theta_W}M_W^2)$
is the Fermi coupling constant, $M_W$ the $W$-boson mass
and $u$ denotes the $u$-quark density $u(x,Q^2)$ etc.

The above formulae apply equally well to muon scattering, whereas for neutrino
scattering some rearrangements are required. For charged current scattering,
eq. (\ref{CCMINUS}) apply for a $\bar{\nu}$ with $\lambda =-1$ and eq.
(\ref{CCPLUS}) for a $\nu$ with $\lambda =+1$, giving the results
in eq. (\ref{CCGF}).
For neutral current neutrino
scattering a similar correspondence applies,
but the electron electroweak
couplings must also be changed to the neutrino ones.
Hence, only the pure $Z^0$ contribution remains
in eqs.\ (\ref{ACOEFF},\ref{BCOEFF}) with $\nu -Z^0$
vector and axial vector couplings, $v_{\nu}$ and $a_{\nu}$ \cite{PDG}.

When simulating the kinematic   variables according to the above differential
cross sections one may choose the two independent variables
depending on the process. The choice $x,Q^2$ for NC and $x,y$ for
CC is suitable and adopted in the program,
but this can be changed (see LST(1) in common LEPTOU)
to other combinations that may improve effiency, e.g., by better reflecting
strong kinematic cuts.
The two variables are first chosen according to a function that can be
directly generated without any rejections, i.e.\ its primitive function
can be obtained and inverted analytically. By chosing this function to
represent the strongest variation of the cross section formulae,
the remaining part can be taken into account by a simple rejection technique
whose efficiency will be higher the smaller the variation in this remaining
function. Formally, a function
\begin{equation}
\label{HV}
  h_v(v) = a_0 + \frac{a_1}{v} + \frac{a_2}{v^2} + \frac{a_3}{v^3}
\end{equation}
is introduced for each variable $v=x,y,Q^2,W^2$ and the cross section
is rewritten in the form (using variables $x,Q^2$ as a definite example)
\begin{equation}
  \label{SIMSIG}
  d\sigma= \left\{ h_x(x)dx \; h_{Q^2}(Q^2)dQ^2 \right\}
           \left\{ \frac{d\sigma / dx dQ^2}{h_x h_{Q^2}} \right\}
\end{equation}
where the random variables can be chosen exactly
according to the expression in the first bracket and the
remaining factor is used for the weighting procedure.
The latter simply means that the chosen $x,Q^2$ point is rejected if
it results in a value for the second factor which is smaller than
a random number (uniformly chosen between zero and unity) times the
maximum of that factor with respect to the two variables ($x,Q^2$ in this case).
Actually, any estimate larger than the maximum will do, but efficiency
improves the closer to the true maximum it is. This estimate is, however, a
constant for any fixed interaction and kinematic   region and can therefore
be obtained in the initialization phase of the Monte Carlo program, see
subroutine LINIT.
Thus new points are generated using the first factor until accepted by
weighting with the second factor and hence the resulting events, or phase
space points, will have no weight associated with them (or more correctly,
they all have unit weight).

The parameters $a_i$ in the functions $h_v$ are in the program represented by
OPT$v(i)$ in common LOPTIM and are given process dependent default values
in subroutine LINIT to optimize program speed under normal conditions.
This means that the functions $h_v$ reflect the dominant variation of the
matrix element from propagators etc. Since this may change with kinematics,
e.g. dominance of pure $Z^0$ exchange at very large $Q^2$, more optimal
values may be found under certain conditions.

In the initialization phase (subroutine LINIT) the simulation
variables are defined and their effective limits calculated from applied cuts.
The optimization parameters are set and the maximum needed for the weighting
is found using an adaptation of {\sc Minuit} \cite{MINUIT}. Further, the total
cross section can be obtained at this stage by numerical integration over
the kinematic   variables. In the simulation phase, see subroutine LEPTO,
phase space points are chosen from the cross section (in subroutine LEPTOX)
as discussed above. This Monte Carlo sampling is also used
to provide an estimate of the cross section for the process being
simulated, see PARL(24) in common LEPTOU.
Since the result is updated with each generated event the accuracy
depends on the generated statistics as
$\sim 1/\sqrt{N}$.

\subsection{Longitudinal structure function}
Target mass effects and the longitudinal structure function, defined by
\begin{equation} \label{FLDEF}
 F_L(x,Q^2) = \left( 1+ 4\frac{x^2m_p^2}{Q^2}\right) F_2(x,Q^2) -
              2xF_1(x,Q^2)
\end{equation}
can be included as an option (see LST(11)) using the formalism for
photon exchange. The cross section in eq. (\ref{SIGMAEM}) is then
modified to \cite{PDG}
\begin{eqnarray} \label{FLSIGMA}
 \frac{d^2\sigma_{\gamma}(ep)}{dx\: dQ^2} & = &
 \frac{4\pi \alpha^2}{xQ^4}\left[
       \left( 1-y-y^2 \frac{x^2m_p^2}{Q^2}\right)
 F_2(x,Q^2) + y^2\, xF_1(x,Q^2) \right] \\
 & = & \frac{2\pi \alpha^2}{xQ^4} \left[
 \left( 1+(1-y)^2 \right) F_2(x,Q^2) -y^2 \tilde{F}_L(x,Q^2) \right]
\end{eqnarray}
where all mass effects have been absorbed into $\tilde{F}_L$ which
consists of the three terms
\begin{equation} \label{FLTERMS}
  \tilde{F}_L(x,Q^2) =
               F_L^{QCD}(x,Q^2) + F_L^{TM}(x,Q^2) + F_L^{HT}(x,Q^2)
\end{equation}
The QCD contribution (which is leading twist) is to order \als \
given by \cite{ROBERTS}
\begin{equation} \label{FLQCD}
  F_L^{QCD}(x,Q^2) = \frac{4\alpha_s(Q^2)}{3\pi}
  x^2 \int_x^1 \frac{dy}{y^3} F_2(y,Q^2)
  + \frac{2\alpha_s(Q^2)}{\pi} \sum_f e_f^2 \: x^2
  \int_x^1 \frac{dy}{y^3} \left(1-\frac{x}{y}\right) yg(y,Q^2)
\end{equation}
The first term originates from the gluon radiation diagram and
the second one from the photon-gluon fusion process, where the sum runs
over the quark flavours (taken as $u,d,s,c$ in the program).
$F_L^{QCD}$ gives an important contribution at small-$x$, where the
gluon term dominates and $F_L$ may hence be used to extract the gluon
distribution \cite{MANDY}.
The target mass correction to ${\cal O}(m_p^2/Q^2)$ is given by
\cite{ROBERTS,HTWIST}
\begin{equation} \label{FLTM}
  F_L^{TM}(x,Q^2) = 4\frac{m_p^2}{Q^2} x^3
  \int_x^1 \frac{dy}{y^2} F_2(y,Q^2)
  - 2 \frac{m_p^2}{Q^2} x^2 F_2(x,Q^2)
\end{equation}
where also the mass term in eq. (\ref{FLSIGMA}) has been included.
Finally the (dynamical) higher twist contribution to
${\cal O}(1/Q^2)$ can be written as \cite{HTWIST,ROBERTS}
\begin{equation} \label{FLHT}
  F_L^{HT}(x,Q^2) = 8\frac{\kappa^2}{Q^2} F_2(x,Q^2)
\end{equation}
where the scale of this twist-4 contribution is given by the parameter
$\kappa^2$ (cf.\ PARL(19)) with a value around 0.03 GeV$^2$
obtained \cite{HTWIST} from SLAC data.

In the numerical implementation of these equations
the evaluation of the integrals in
eqs. (\ref{FLQCD},\ref{FLTM}) can be performed for each event at its
proper $x$ and $Q^2$-value. Since this tends to be time-consuming
another option (see LST(11)) is to initially (in LINIT) set up a grid 
in $x,Q^2$ with their
values and then perform a linear interpolation to the desired $x,Q^2$
when simulating events. The latter method is preferrable and
gives almost the same precision under normal conditions.
The inclusion of $F_L$ only affects the distribution in $x,Q^2$, i.e.\
eq. (\ref{FLSIGMA}) and thereby the cross section, but not the
generation of the hadronic part of the event. For the first order QCD 
matrix element corrections, which are still generated as usual 
(cf.\ section 2.5), this is not fully consistent but should be an 
adequate approximation for inclusive properties of the hadronic final state.

\subsection{Parton density distributions}
\label{sec:parton-density}
To define the parton content of the proton for the cross section
formulae above the parton density functions $q_f(x,Q^2),\,
\bar{q}_f(x,Q^2)$ and $g(x,Q^2)$ are needed. These give the probability
to find a quark or antiquark of flavour $f$, or a gluon, respectively,
carrying a fraction $x$ of the proton momentum when probing the proton
with a momentum transfer $Q^2$.
Several parametrizations of these distributions have been obtained using
data, in particular from lepton scattering experiments, and with
$Q^2$-dependence according to the perturbative QCD evolution equations
\cite{AP}. 
The fit to the data provides the $x$-dependence and the QCD parameter
$\Lambda$. The choice among many available parton density parametrisations 
in {\sc Pythia} 5.7 \cite{JETSET} and in {\sc Pdflib} \cite{PDFLIB} 
is made through the switches LST(15) and LST(16) in common LEPTOU.

The possibility to scatter on intrinsic charm and bottom quarks in the 
nucleon is included by an option, see LST(15). 
The hypothesis of intrinsic charm quarks in the proton was introduced 
\cite{Brodsky80} as an attempt to understand a large discrepancy
between  early charm hadroproduction data and leading order
perturbative QCD (pQCD) calculations. This discrepance has 
largely disappeared as more data have been collected and next-to-leading  
order (NLO) pQCD calculations have been made. Still, however, there are 
certain aspects of charm production data which are difficult to understand
within the pQCD framework, but are  natural if the intrinsic charm
hypothesis is basically correct (see \cite{IC} and references therein).
Intrinsic charm (IC) corresponds to a Fock-state
decomposition of the proton wave function,  $|p\rangle = \alpha
|uud\rangle + \beta |uudc\bar{c}\rangle +...$,  with a small, but
finite, probability $\beta^2$ (PARL(12)) for an intrinsic $c\bar{c}$ pair 
as a quantum fluctuation.
The normalization of this component is the key unknown, 
although it should decrease as $1/m_Q^2$.  Originally, a 1\% \ probability 
was assumed, but later investigations based on EMC data on the charm 
structure function $F_2^c(x,Q^2)$ \cite{Aubert82} 
indicate a somewhat smaller but non-vanishing level; 
$\sim 0.3\%$ \cite{Hoffmann} and $(0.86\pm 0.60)\%$ \cite{Harris}. 

From the IC model one obtaines an effective charm quark density \cite{IC}
\begin{equation}\label{eq:c(x)}
c_{IC}(x)=\beta^2 1800x^2\left\{
\frac{1}{3}(1-x)(1+10x+x^2)+2x(1+x)\ln{x}\, \right\}
\end{equation}
which gives a charactersitic hard momentum spectrum 
with $\langle x_c \rangle =2/7$.
The $Q^2$ dependence from normal leading log GLAP equations have been 
calculated for IC \cite{Hoffmann}, but can be taken into account through a  
simple extension of the parameterisation in Eq.~(\ref{eq:c(x)}) \cite{IC}.
This quark density is then used in the electroweak cross section fomulae
to simulate the scattering on such an intrinsic charm (or bottom) 
quark, with parton showers and hadronization added as usual. 
The scattered (anti)charm quark gives a charmed hadron in the current
fragmentation region and the remaining partner (anti)charm quark in 
the proton remnant gives a charmed hadron in the target fragmentation
region. For phenomenological studies with this model, see \cite{IC}.

\subsection{First order QCD processes}
\label{sec:ME}
The leading order parton level process is $V^{\star}q\to q$, where
$V^{\star}$ is the exchanged virtual boson $\gamma /Z$ or $W$.
In first order QCD the gluon radiation or QCD Compton process, 
$V^{\star}q\to qg$, and the boson-gluon fusion (BGF) process, 
$V^{\star}g\to q\bar{q}$, appear and can
be included with their matrix elements (ME) \cite{AM,PR,QCDME,mixed}.
Quark masses are not included in these ME, but a threshold factor is
applied for boson-gluon fusion into heavy quarks. For more accurate
heavy quark simulations the AROMA Monte Carlo \cite{AROMA} may be used.

The first order ME are rather complicated functions involving three new degrees
of freedom corresponding to energy, polar angle and azimuthal angle of one
final parton (the other is then determined by energy-momentum conservation). In
terms of the more suitable variables \cite{PR} $x_p=x/\xi$ and $z_q=P\cdot
p_q/P\cdot q$, where $\xi$ is the momentum fraction that the incoming parton
take of the proton and $p_q$ the four-momentum of,  e.g., the final quark, the
cross sections are five-fold differential
\begin{equation} \label{SIGMAQCD}
  \frac{d^5 \sigma}{dx\, dQ^2 \: dx_p\, dz_q\, d\phi}
\end{equation}
Here $\phi$ is the parton azimuthal angle with respect to the lepton scattering
plane (in the $\gamma p$ cms) and its distribution has the form
\begin{equation} \label{SIGMAPHI}
  d\sigma = d\sigma_0 + d\sigma_1 \cos{\phi} + d\sigma_2 \cos2{\phi}
\end{equation}
where $d\sigma_i$ depends on the other four variables. When averaging over
$\phi$, only the first term contributes and the $\phi$-dependence can therefore
often be neglected, but in dedicated analyses it can be observed
\cite{HEDBERG}. The invariant mass squared of the two emerging partons 
($qg$ or $q\bar{q}$) is labelled $\hat{s}$.

The matrix elements have soft and collinear divergences that may be partly
cancelled by virtual corrections and partly absorbed in the parton density 
functions. The QCD-Compton cross section $\hat{\sigma}_{qg}$
diverges as $1/(1-x_p)(1-z_q)$ and the boson gluon fusion 
$\hat{\sigma}_{q\bar{q}}$ as $1/z_q(1-z_q)$. 
To avoid these singularities in a Monte Carlo simulation procedure, 
it is common practice to impose a cut-off on the matrix elements. 
In {\sc Lepto}, several different cut-off procedures are available 
through LST(20). 
In the $W$ scheme it is required that each pair of emerging partons 
(including the proton remnant) have 
a minimum invariant mass, expressed as $s_{ij}=(p_i+p_j)^2>y_{cut}W^2$.
As a new default in the present version of the program, we use 
a variation of the `mixed scheme' \cite{mixed}. 
In this `$z\hat{s}$' scheme the applied cuts are 
$z_{q,min}<z_q<1-z_{q,min}$ and $\hat{s}>\hat{s}_{min}$, with 
$z_{q,min}$ and $\hat{s}_{min}$ set by PARL(8) and PARL(9), respectively. 
As discussed in \cite{Unified}, this gives some advantages compared to
previously used cut-off schemes. In particular, it gives a better division 
of the available phase space into a region for hard emission, which is best
described by exact matrix elements, and the soft and collinear regions where
higher orders are important and can be taken into account by leading log 
parton showers (cf.\ section 2.6). 

To decide on an event-by-event basis whether to generate one of the first order
event types ($qg$- or $q\bar{q}$-event) rather than the leading order process
($q$-event), the probability for each event type must be available as a
function of the kinematic variables $x,Q^2$. The $q$-event probability is taken
as $P_q=1-P_{qg}-P_{q\bar{q}}$ and the probabilities $P_{qg}$ and
$P_{q\bar{q}}$ are defined as the integral of the relevant first order matrix
element over the above three variables, divided by the overall differential
cross section $d\sigma /dxdQ^2$. The integration over $\phi$ is trivial and the
variable $z_q$ can be integrated analytically leaving only the
$x_p$-integration, which involves parton density parametrizations, to be
performed numerically using an adaptive Gaussian method.  If the cut-off is
chosen too low the calculated probabilities for the first order processeses
$P_{qg}+P_{q\bar{q}}$ can be larger than unity at the given $x$ and $Q^2$. In
this case the cutoff is increased, the $y_{cut}$ in the $W$ scheme and the
$z_{min}$ or the $\hat{s}_{min}$ in the $z\hat{s}$ scheme, until
$1-$PARL(13)$<P_{qg}+P_{q\bar{q}}<1$. The effective cut-off actually used in
a generated event is stored in PARL(27).

To save computer time the probabilities for $qg$- and $q\bar{q}$-events can be
calculated at the initialization time (LINIT) and stored on a `grid' in the
$x,W$ or $x,y$ plane. In the following event generation phase the probabilities
at any $x,Q^2$ (or $x,y$) point, chosen according to section 2.2, are then
obtained by linear interpolation on this grid. The switch LST(19) regulates the
use of such grids and provides an option with a grid that automatically 
adjusts to the kinematical region chosen by the user. There is also a
possibility to calculate these integrals for each event, and thereby avoid
interpolation errors. 

Having chosen to generate a first order event based on the probabilities $P_q$,
$P_{qg}$ and $P_{q\bar{q}}$, the internal variables $x_p,z_q$ and $\phi$ are
generated in turn from the matrix element formulae where the subsequent
variables are integrated out. Given the values of all five variables the
four-momenta of the scattered lepton and partons can be calculated. 

\subsection{QCD parton shower evolution}
In order to take higher than first order QCD effects into account the parton
shower (PS) approach has been implemented as described in detail in ref.\
\cite{PS}. This has the advantage that arbitrarily high orders in \als \ can be
simulated, but only in the leading $\log Q^2$ approximation as opposed to the
exact treatment in fixed order ME. Higher order effects are important at 
high energies where multiple parton emission can give rise to multijet events 
as well as affect the internal properties, such as hardness and
width, of a jet \cite{BIS,BIR} and the overall structure of the event
in terms of, {\em e.g.}, energy flows.

In DIS the quark struck by the electro\-weak boson can emit partons both before
and after the boson vertex giving rise to initial and final state parton
showers, respectively. A parton close to mass-shell in the incoming nucleon can
initiate a parton emission cascade (or shower) where in each branching one
parton becomes increasingly off-shell with a space-like virtuality ($m^2<0$)
and the other is on-shell or has a time-like virtuality ($m^2>0$). This initial
state space-like shower results in a space-like quark which interacts with the
electro\-weak boson that turns it into an outgoing quark which is either
on-shell or has a time-like virtuality. In the latter case a final state,
time-like shower will result where the off-shell mass is reduced by branching
into daughter partons with decreasing off-shell masses and decreasing opening
angles. This shower continues until all partons are (essentially) on-shell. Any
parton with a time-like virtuality from the initial state shower will develop
similarly. The general behaviour of initial and final state showers are similar
since they are both based on the branching processes $q\to qg$, $g\to gg$ and
$g\to q\bar{q}$ as described by the GLAP equations \cite{AP} in the
leading $log Q^2$ approximation of perturbative QCD.

The final state radiation is analogous to parton radiation in $e^+e^- \to
q\bar{q}$ and is theoretically well developed and tested against data. The
routine LUSHOW in {\sc Jetset} \cite{JETSET} is therefore used for all
time-like showers. The evolution is based on the Sudakov form factor, which
expresses the probability that a parton does not branch between some initial
maximum virtuality and some minimum value. From this one can find the mass of
the decaying parton, the energy fractions in the branching and the flavours of
the daughter partons. The process is iterated with a reduced virtuality until
all parton virtualities are below some cutoff $m_0^2$ around 1 GeV$^2$. The
technical details are given in \cite{PS}, but it should be noted that coherence
in soft gluon emission is taken into account through angular ordering
(decreasing opening angles in subsequent branches) and that $p^2_\bot \simeq
z(1-z)m^2$ is used as argument in \als \ as suggested by studies of coherence
effects.

The initial state radiation is performed using the `backwards' evolution scheme
\cite{BACKWARD} where the shower is constructed from the hard electroweak
interaction backwards with decreasing virtualities down to the on-shell parton
from the incoming nucleon. A modified version \cite{PS} of the routine PYSSPA in
{\sc Pythia}  \cite{PYTHIA48} is used for the space-like shower. This is a more
complicated process, e.g. since the nucleon parton density functions must be
taken into account (which tend to reduce the amount of radiation). In addition
it is not so well tested by the more `messy' hadron collision data. When
combining the initial and final state radiation to get the complete model,
special precautions have been taken \cite{PS} to preserve energy-momentum
conservation and keep the normal definitions of the kinematic variables for the
electroweak scattering such that they will be obtained from the scattered
lepton as usual, in particular that Bjorken-$x$ is preserved.

The amount and hardness of the initial and final radiation depends on the
off-shellness of the struck parton just before and the partons after the boson
vertex. These virtualities are chosen, using the Sudakov form factor, between
the lower cut-off and a maximum value to be given by the energy or momentum
transfer scale in the process. The default procedure is to add the parton
shower to a matrix element event. The possible final states are then $q$, $qg$
and $q\bar{q}$ events. 
In the case of a $q$ event the maximum scales are set by the matrix
element cut-off (e.g. $y_{cut}W^2$ in the $W$-scheme) 
since emissions harder than the cut-off would be double-counting.
In the case of a $qg$ or $q\bar{q}$ event the maximum virtuality scale 
for the final state shower is set to $\hat{s}$. 
As maximum virtuality for the backwards initial state shower it is natural
to use the mass-squared of the quark propagator just before the boson vertex.
This is given by the known four-vectors of the exchanged boson and the two 
final partons from the matrix element, 
but depending on the underlying Feynman diagram in the amplitude
different combinations are possible leaving some remaining ambiguity. 
The largest of these possible virtualities is being used (cf.\ subroutine
LSCALE). 

In the case where the parton shower is used without the matrix element there is
a large ambiguity in which maximum scale should be used. Whereas only one scale
is present in $e^+e^-$, any function of $Q^2$ and $W^2$ may be possible in
DIS. Although $Q^2$ and $W^2$ can often be of similar magnitude, cf.\ eq. (2),
at small $x$-values $W^2$ is much larger than $Q^2$ giving rise to drastically
different amounts of radiation \cite{CASCADE}. Given this uncertainty,
different scales are available (cf.\ LST(9)) based on different motivations.
The phase space limit is given by $W^2$, such that a smaller scale ($Q^2$)
would cut off the tail of high-$p_\bot$ parton emission.  Refering to the
first order ME parton level one finds $\langle p^2_\bot \rangle \sim Q^2(1-x)$
for $x\to 1$ and $\langle p^2_\bot \rangle \sim Q^2ln(1/x)$ for $x\to 0$
\cite{AM}. A suitable interpolation between these limiting behaviours is the
choice $Q^2(1-x)max(1,\ln{\frac{1}{x}})$ which is the default choice when the
matrix elements are not used.

The parton shower approach has some shortcomings due to its approximate nature.
The separation of initial and final state parton emission implies the neglect
of interference terms between the two and is not gauge invariant. The use of
the leading logarithm approximation means that the emission of partons that are
soft or close to the directions of the emitting partons should be well
described, while the emission of hard partons at large angles could be
mistreated. Therefore, the rate of events with extra hard partons that give
rise to separate jets, i.e.\ multijet events, need not be well accounted for.
The use of matrix elements is preferable for these purposes and is therefore
the default option. Starting with the ME the hard emission is generated and
extra, but softer emissions are then added using PS. An advantage of this
procedure is, of course, that the hard parton emission is properly treated
using the first order ME.

\subsection{Nucleon remnant and hadronization}
The remnant system is the target nucleon `minus' the parton entering 
the hard scattering system (initial parton showers and matrix
elements). This interacting parton can be either a valence quark, 
a sea-quark or a gluon.

When the interacting parton is a valence quark the nucleon remnant is simply a
diquark composed of the two left-over valence quarks as spectators. In the Lund
model \cite{LUND} a colour triplet string is stretched between the colour
triplet charged struck quark and the diquark which is a colour antitriplet.
This system is then hadronized in the usual way \cite{LUND,JETSET} by the
production of quark-antiquark and diquark-antidiquark pairs from the energy in
the  field,  leading to hadron production. The proton remnant diquark is not a
single entity; its two quarks may go into a leading baryon but they can also be
separated to produce a leading meson followed by a baryon.

In case the interacting parton is a sea quark ($q_s$) or antiquark  the
nucleon remnant contains the corresponding antiquark or quark  in
addition to the three valence quarks ($q_v$). This more complicated
four-quark system $q_vq_vq_v\bar{q}_s$ or $q_vq_vq_vq_s$  
must be taken into account to conserve the flavour quantum numbers. 

In the conventional way (default in {\sc Lepto} version 6.2 and
earlier) the following treatment has been used. If $\bar{q}_s=\bar{u}$
or $\bar{d}$ it is cancelled against a corresponding valence quark
leaving a simple diquark system to be treated as above. For other flavours of
$\bar{q}_s$ it is joined with a valence quark of arbitrary flavour into
a meson ($M=q_v\bar{q}_s$). The $\bar{q}_s$ is assumed to have no
specific dynamic properties such that this splitting process into a
meson and a diquark should be similar to normal hadronization. The
meson is then given a fraction $z$ of the remnants energy-momentum
($E+p_z$) along the beam direction from a probability distribution
$P(z)$ (cf.\ LST(14)) and only a small Gaussian $p_\bot$ (cf.\
PARL(14)). The left-over diquark, with longitudinal momentum given by
$1-z$ and equal but opposite $p_\bot$, forms a string system with the
scattered quark and hadronization proceeds as usual. If an antiquark
($\bar{q}_s$) was scattered the remnant is a four-quark system
$q_vq_vq_vq_s$ which is treated similarly to the previous case. Here,
the corresponding quark ($q_s$) is combined with a random diquark
giving a baryon ($B=q_vq_vq_s$) leaving the remaining valence quark to
form a string system with the scattered antiquark. The split of the
remnant is as before, taking account of the masses in the distribution
for $z$ (cf.\ LST(14)).

In {\sc Lepto} 6.3 a modified treatment of sea quarks in the remnant 
was introduced which is now default (cf.\ LST(35)). The essential
difference is that the sea quark partners ($\bar{q}_s$) are treated 
dynamically and also $u$ and $d$ quarks can be considered as sea quarks.
The interacting quark is assigned to be a valence or sea
quark from the relative size of the corresponding parton distributions
$q_v(x_1,Q^2_1)$ and $q_s(x_1,Q^2_1)$, where $x_1$ is the momentum
fraction of the quark `leaving' the proton and $Q^2_1$ is the relevant
scale (typically the cutoff $Q^2_0$ of the initial state parton
shower). In case of a valence quark the previous treatment is used, but
in case of a sea quark a new treatment is used. The left-over partner
$\bar{q}_s$ is given a longitudinal momentum fraction from the
Altarelli-Parisi splitting function $P(g\to q\bar{q})$ and the
transverse momentum follows from the masses of the partons in the
splitting. Essentially the same results are obtained if the
longitudinal momentum fraction is chosen from the corresponding sea quark 
momentum distribution. The former approach is presently used since this
allows the mechanism to be simply implemented in the initial state
parton shower routine as an additional, but non-perturbative, $g\to
q\bar{q}$ process. This partner sea quark will then be at the end-point
of a string and not, as previously, go directly into a hadron together
with another spectator parton. Depending on the momentum of the partner
sea quark, this new string may extend more or less into the central
region and through hadronization contribute to the particle and energy
flow in the forward region. In particular, the transverse forward
energy flow will be enhanced \cite{SCI,Unified} and improve the agreement with
HERA data. 

In boson-gluon fusion the removed gluon leaves the three valence quarks
in a colour octet state. This remnant is split into a quark
and a diquark, chosen with random flavours, which form two separate
strings with the antiquark and quark, respectively, produced in the
fusion process. Again the split of the remnant involves the same
longitudinal momentum sharing and a Gaussian transverse momentum.
For the order \als \ gluon radiation process ($qg$-event) the string
is stretched from the scattered quark via the gluon to the target
remnant.

In the parton shower case, the backwards evolution always results in
one parton being removed from the nucleon as in
the above cases such that the same procedures can be applied.
The additional partons emitted in the PS case will, however, lead to
a more complicated string configuration.
The string follows the colour flow of the parton shower such that it
starts from a colour triplet quark and goes via a number of colour
octet gluons, which are kinks on the string, before ending up on
a colour antitriplet antiquark or diquark. Where quark-antiquark
pairs have been produced in the shower, the colour flow will be
broken resulting in a termination of the first string piece and the
start of a new one. The string system may thus be divided into subunits
which then hadronize separately.

The ME and PS emissions may give a varying number of soft or 
collinear partons, depending on the details of the cut-offs.
Although such partons cannot be observed as separate jets, 
they may give a `softening' and `fattening' of jets. 
The Lund string model is particularly suitable in this context, 
since it provides a stability in the sense that the hadron level 
result will not depend strongly on the presence of extra soft partons. 
Rather, one obtains a smooth transition to a configuration without them 
\cite{SMOOTH1,SMOOTH2}. The independent hadronization model, available 
as an option in \cite{JETSET} does not have the same property and is 
therefore not recommendable.

In this context one should also note that the two-string configuration 
for sea-quark initiated processes provides a desirable continuity between
the two-string gluon-initiated $q\bar{q}$-events and the one-string 
quark-initiated $q$-events. Depending on the partner sea-quark momentum, 
the corresponding string will extend more or less into the central 
region in rapidity. The hadronization of this extra string will contribute 
to the particle multiplicity and energy flow in this region \cite{SCI,Unified}.

The parameters for the hadronization process in {\sc Jetset}
\cite{JETSET} are obtained from fits to $e^+e^-$ data and are assumed
to be the same in DIS based on jet universality. Nevertheless, they
depend on which QCD effects are explicitly included in the Monte Carlo
simulation. The default values are suitable when higher orders are
taken into account via parton showers, whereas with first order ME alone 
the hadronization should be made slightly `softer' and `wider' to account
for the additional parton emission not simulated explicitly.

\subsection{Soft colour interactions and rapidity gaps} 

The rapidity gap events discovered in deep inelastic scattering at HERA 
\cite{HERA-gaps} are usually interpreted in terms of pomeron exchange models
\cite{Pomeron}. Although this seems to work reasonably well phenomenologically,
there is no satisfactory understanding of the pomeron and its interactions 
mechanism. As an alternative, we have introduced a model \cite{SCI,Unified}
based on soft colour interactions (SCI) that give rise to rapidity gap events
without using the concept of a pomeron. 

At small Bjorken-$x$ ($10^{-4}-10^{-2}$), where the rapidity gap events occur,
the events are frequently initiated by a gluon from the proton. This can either
be directly from the boson gluon fusion matrix element or after the initial
state parton shower, including a possible split in the sea quark treatment. In
the conventional string hadronization model this gives two separate strings
from the $q$ and $\bar{q}$ to the proton remnant spectator partons, where the
gluons from the parton shower are kinks on the string, thereby causing
particle production over the whole rapidity region in between. The new
hypothesis introduced here is that additional non-perturbative soft colour
interactions may occur. These have small momentum  transfers, below the scale
$Q_0^2$ defining the limit of pQCD, and do not significantly change momenta
from the perturbative phase. However, SCI will change the colour of the partons
involved and thereby change the colour topology as represented by the strings.
Thus, it is proposed \cite{SCI} that the perturbatively produced quarks and
gluons can interact softly with the colour medium of the proton as they
propagate through it. This should be a natural part of the processes in which
`bare' perturbative partons are `dressed' into non-pertubative quarks and
gluons and the formation of the confining colour flux tube in between them. 

Lacking a proper understanding of such non-perturbative QCD processes, 
a simple model is used to describe and simulate these interactions. 
All partons from the hard interaction (electroweak $+$ pQCD) plus the 
remaining quarks in the proton remnant constitute a set of colour
charges. Each pair of charges can make a soft interaction changing only
the colour and not the momenta, which may be viewed as soft
non-perturbative gluon exchange. As the process is non-perturbative the
exchange probability cannot be calculated so instead it is described by
a parameter, PARL(7). The number of soft exchanges will vary
event-by-event and change the colour topology of the events such that,
in some cases, colour singlet subsystems arise separated in rapidity.
In the Lund model this corresponds to a modified string stretching and 
rapidity gaps may arise when a gap at the parton level is not spanned by 
a string, as illustrated in \cite{SCI}. 
In particular, when the hard process starts with a gluon from the proton, 
leaving a colour octet remnant and giving a colour octet hard scattering 
system, a soft colour exchange between the two octet systems can give two 
colour singlets separated in rapidity. 

SCI may thereby give a string system, including a valence diquark from the
proton remnant, which has a small invariant mass. Such systems are not
optimally treated in {\sc Jetset} regarding production of one- or two-particle
final states and taking isospin constraints into account. We have therefore
constructed a new treatment (implemented in subroutine LSMALL) for such
systems giving two particles if kinematically possible and otherwise one. When
one particle is made, the isospin restriction from the effective isospin
singlet exchange in SCI is taken into account to prevent $\Delta$ production.
The invariant mass of such small systems is roughly given by the constituent
masses of the partons in the proton remnant and the transverse momentum in the
remnant split. With the default contituent masses ($m_q=325$ MeV and
$m_{qq}=650$ MeV) and width of the transverse momentum in the remnant split
($350$ MeV), the invariant mass of the consituents in the proton remnant
exceeds the mass of the proton itself. Therefore the possibility has been
introduced to reduce the constituent masses of the remnant partons when the
target remnant is more complicated than a simple diquark (cf.~PARL(20)).

SCI also gives rise to other string topologies without gaps, but where 
the string goes back and forth in rapidity between the partons. In such cases,
there will be more particles and energy per unit of rapidity. 
This contributes to a better description of, {\em e.g.}, the forward 
energy flow observed at HERA as discussed in \cite{Unified}. 
  

\section{Description of program components}
The present program, {\sc Lepto} version 6.5, is a backwards compatible update 
of versions 6.1 \cite{LEPTO61} to 6.4. It is a further development of
earlier versions \cite{LEPTO}, but is not backwards compatible
with those due to the changed conventions in {\sc Jetset} version 7 now
being used.
Previous knowledge of earlier versions is not necessary, but is helpful
since the main structure and use of the program has been kept similar.
The code is written completely in standard FORTRAN77 and should
therefore run on any computer with such a compiler.
Single precision is normally used, but double precision is being used
when required.
\subsection{Subroutines and functions}
The following routines should or may be called by the user:

\noindent
SUBROUTINE {\bf LINIT}(LFILE,LEPIN,PLZ,PPZ,INTER)
\begin{defl}{123456789012}
\item[{\it Purpose:}]
to initialize the event generation procedure and, optionally, integrate
cross section.
\item[{\it Arguments:}]
\item[LFILE~:]
logical file number containing weights for first order QCD, see LST(8).
\item[{\hfill $=0$:}]
the weights are calculated but not save on file, no file is used.
\item[{\hfill $<0$:}]
the weights are calculated and stored on file number --LFILE.
\item[{\hfill $>0$:}] the weights are read from file number LFILE.
\item[{\hfill {\it Remarks:}}]
When using weights from a file, the same conditions (interaction, cuts
etc) must hold as when the weights were calculated.
The relevant quantities are checked and a mismatch results in an error
message, cf.\ LST(3).
There is no strong reason, in this upgraded program, to store weights on
a file since their calculation is fast enough to be repeated in each
run.
\item[LEPIN~:] type of lepton,
i.e.\ $11=e^-,\: 12=\nu_e,\: 13=\mu^-,\: 14=\nu_{\mu}$ and negative
values for the coresponding antiparticles, i.e.\ {\sc Jetset} \cite{JETSET}
code.
\item[PLZ, PPZ~:]
momentum (GeV/c) for incoming lepton and nucleon, respectively,
along the $z$-axis (if both non-zero, i.e.\ colliding beams, they must
have opposite signs).
Colliding beams which are not along a common axis, or variable beam
energies, are possible using LST(17).
\item[INTER~:]
type of interaction to be simulated.
\item[{\hfill =1:}]  electromagnetic (EM), i.e.\ $\gamma$ exchange.
\item[{\hfill =2:}]  weak charged current (CC), i.e. $W^{\pm}$ exchange.
\item[{\hfill =3:}]  weak neutral current, i.e.\ $Z^0$ exchange.
\item[{\hfill =4:}]  neutral current (NC), i.e.\ $\gamma /Z^0$ exchange.
\item[{\it Procedure:}]
Various constants are calculated, effective
limits on kinematic   variables are derived from the cuts and,
depending on the interaction,
integration variables are chosen and parameters are set.
For the Monte Carlo rejection technique to give unweighted events, the
maximum of the differential cross section (having factored out a
suitable function that can be simulated exactly, see section 2.2) is
found using an adaptation of {\sc Minuit} \cite{MINUIT}.
The cross section is calculated using numerical integration (LST(10))
and stored (PARL(23)).
Optionally, grids with probabilities for first order QCD processes are
set up (LST(8)) and grids for including the longitudinal
structure function (LST(11)).
\end{defl}

\noindent
SUBROUTINE {\bf LEPTO}
\begin{defl}{123456789012}
\item[{\it Purpose:}]
to administer the generation of one event of the kind specified by the last
LINIT call.
\item[{\it Procedure:}]
Beam energies are taken from LINIT (optionally from LUJETS, see
LST(17)).
A phase space point $x,Q^2$ is chosen according to the differential
cross section (using LEPTOX), and the parton level system, optionally
with QCD corrections (cf.\ LST(8)) is set up.
Hadronization and decays are performed, via LUEXEC, and the event is
transformed to the selected frame (LST(5),LST(6)).
The Monte Carlo estimate of the cross section, PARL(24),
is updated with each event.
\item[{\it Remarks:}]
Under some conditions, an error may have occured and the
event should be rejected, see LST(21).
\end{defl}

\noindent
SUBROUTINE {\bf LFRAME}(IFRAME,IPHI)
\begin{defl}{123456789012}
\item[{\it Purpose:}]  to transform the event between different frames.
\item[{\it Arguments:}]
\item[IFRAME~:] specifies the desired frame (as for LST(5)).
\item[{\hfill =1:}] hadronic CM frame, $z$-axis along exchanged boson.
\item[{\hfill =2:}] lepton-nucleon CM frame, $z$-axis along lepton.
\item[{\hfill =3:}] lab system as defined by last user call to LINIT.
\item[{\hfill =4:}] as 3, but $z$-axis along exchanged boson.
\item[IPHI~:]  specifies whether to include a random rotation for the
     azimuthal angle, $\phi$, of the lepton scattering plane.
\item[{\hfill =0:}] no rotation, scattering plane is $x-z$ plane.
     A possible earlier rotation is undone.
\item[{\hfill =1:}] random rotation in $\phi$, made
     in lepton-nucleon CM frame. Not made for IFRAME=1.
\item[{\it Remark:}]
     The present frame is stored in LST(28), LST(29) and is updated
     by LFRAME. Transforming the event with other routines can
     therefore cause errors in a following LFRAME call.
\end{defl}

\noindent
SUBROUTINE {\bf LPRWTS}(NSTEP)
\begin{defl}{123456789012}
\item[{\it Purpose:}]
     to print a table of the QCD weights in common LGRID,
     i.e.\ the probabilities for $q$-, $qg$- and $q\bar{q}$-events
     in first order QCD.
     Only the values on each NSTEP point in the $x,W$ grid is printed,
     i.e.\ NSTEP=1 prints all grid points.
\end{defl}

\noindent
SUBROUTINE {\bf LWBB}(ENU)
\begin{defl}{123456789012}
\item[{\it Purpose:}]
to give energy (ENU in GeV) of a (anti-)neutrino chosen from a simple
parametrization (defined by DATA statement within the routine) of a wide
band beam. The energy is actually chosen from the beam spectrum weighted
with the energy to take into account the linear rise with energy of
the cross section, which {\sc Lepto} does not account for.
This routine is a simple example to be
replaced by a more realistic beam energy distribution if necessary.
\item[{\it Remark:}]
For running with variable beam energies, see LST(17).
\end{defl}

\noindent
SUBROUTINE {\bf LTIMEX}(TIME)
\begin{defl}{123456789012}
\item[{\it Purpose:}]
     to get cpu execution time (in seconds) since start of job.
     Interface to machine dependent routine, by
     default  TIMEX (Z007 in the CERN library) which can
     simply be changed, or TIME set to zero since timing information is
     not essential (although useful).
\end{defl}

\noindent
SUBROUTINE {\bf LNSTRF}(X,Q2,XPQ)
\begin{defl}{123456789012}
\item[{\it Purpose:}]
     to give the parton distribution functions per nucleon for a
     nucleus defined by PARL(1) and PARL(2). Arguments as in LYSTFU
     which is called. This routine is called internally, but
     can also be used separately after initialization by LINIT.
\end{defl}

\noindent
SUBROUTINE {\bf LYSTFU}(KF,X,Q2,XPQ)
\begin{defl}{123456789012}
\item[{\it Purpose:}]
     to give the parton distribution functions for the target hadron 
     through an interface to subroutine PYSTFU in {\sc Pythia} 5.7 
     \cite{JETSET}, complemented with options for intrinsic charm and
     beauty quarks in the nucleon.
\item[{\it Arguments:}]
\item[KF~:]  particle flavour, e.g. 
       $2212$=$p$, $2112$=$n$, $-2212$=$\bar{p}$, $-2112$=$\bar{n}$.
\item[X~~:]  momentum fraction carried by the parton.
\item[Q2~:]  momentum transfer scale $Q^2$.
\item[XPQ :]
       array ($-6$:6) that on return contains the values of the
       momentum-weighted parton densities, i.e.\
       $x\cdot q(x,Q^2)$ and $x\cdot g(x,Q^2)$.
       Index: 0=$g$, 1=$d$, 2=$u$, 3=$s$, 4=$c$, 5=$b$, 6=$t$ and
       $-1$=$\bar{d}$, $-2$=$\bar{u}$, $-3$=$\bar{s}$, $-4$=$\bar{c}$,
       $-5$=$\bar{b}$, $-6$=$\bar{t}$.
\item[{\it Remarks:}]
     Different parametrizations of structure functions are implemented
     and can be selected by LST(15) and LST(16).
     This routine is called internally but
     can also be used separately after initialization by LINIT.
\end{defl}

In the following list all subroutines (S) and functions (F) are briefly
described. The order is as they appear in the code and reflects the
flow in the program. Further details about their purpose and procedures
used are given by comments in the code.
All routine names start with characteristic letters to indicate origin and
avoid name clashes. L, or D for real functions, is for {\sc Lepto} routines in
general; FL for routines related to longitudinal structure function;
LY for modified {\sc Pythia} routines \cite{PYTHIA48};
LM for modified {\sc Minuit} routines \cite{MINUIT}; GADAP, RIW and DV for
routines related to different integration procedures.

\noindent
\begin{defl}{123456789012}
\item[{\it Routine}]  {\it Purpose}
\item[LTIMEX \hfill (S)]
     to give execution time since start of job, see above.
\item[LEPTOD  \hfill ~~~]
     block data to give default values to all switches and parameters.
\item[LINIT   \hfill (S)]
     to initialize program package, see above.
\item[LEPTO   \hfill (S)]
     to administer the generation of an event, see above.
\item[LEPTOX  \hfill (S)]
     called by LEPTO to generate kinematic variables, within the
     applied cuts, according to the differential cross section giving
     unweighted events.
     Optionally include $F_L$.
     Update Monte Carlo estimate of cross section, PARL(24),
     select lepton helicity and target nucleon.
\item[LKINEM  \hfill (F)]
     called by LEPTOX to calculate various kinematic variables
     and optionally (LST(2)) to reject event if outside given kinematic
     limits in CUT.
\item[LQCDPR  \hfill (S)]
     called by LEPTO to obtain probabilities for first order QCD processes,
     either by linear interpolation of weights stored on a grid 
     set up in LINIT, or direct calculation event-by-event, see LST(19).
\item[LQEV    \hfill (S)]
     called from LEPTO to generate parton system for $q$-event,
     i.e.\ without QCD processes.
\item[LQEVAR  \hfill ~~~]
     entry in LQEV used by {\sc Ariadne} \cite{ARIADNE}
     for some boson-gluon fusion events.
\item[LQGEV    \hfill (S)]
     called from LEPTO to generate parton system for $qg$-event,
     i.e.\ first order QCD gluon radiation.
\item[LQQBEV   \hfill (S)]
     called from LEPTO to generate parton system for $q\bar{q}$-event,
     i.e.\ first order QCD boson-gluon fusion.
\item[LXP      \hfill (S)]
     called from LQGEV and LQQBEV to generate a value of $x_p$ from the
     relevant QCD matrix element folded with parton density functions
     (for given $x,Q^2$, but $z_q$ and $\phi$
     integrated out).
\item[LZP      \hfill (S)]
     called from LQGEV and LQQBEV to generate a value of $z_q$,
     from QCD matrix element (for given $x,Q^2,x_p$, but $\phi$ integrated out).
\item[LQMCUT   \hfill (F)]
     to apply cuts on QCD parton configuration to take quark mass effects
     into account and ensure ability to apply string hadronization.
\item[LAZIMU   \hfill (S)]
     called from LQGEV and LQQBEV to generate the azimuthal angle $\phi$,
     of parton plane w.r.t.\ lepton scattering plane, from first order
     QCD matrix elements eq.\ (\ref{SIGMAPHI}), for given $x,Q^2,x_p,z_q$.
\item[DSIGMA   \hfill (F)]
     differential cross section $d\sigma /dx_p$ for first order QCD
     processes.
\item[DSIGM2   \hfill (F)]
     modified DSIGMA with a variable substitution to speed up integration
\item[DQCD     \hfill (F)]
     first order QCD differential cross sections
     $d\sigma /dx_p dz_q$ (section 2.5) \cite{PR}.
\item[DQCDI    \hfill (F)]
     first order QCD differential cross section $d\sigma /dx_p$ obtained
     after analytical integration over $z_p$ and factoring out $1/(1-x_p)$.
\item[LFLAV    \hfill (S)]
     selects flavour of struck quark and outgoing quark (with flavour
     mixing in CC) and defines the corresponding parton system from the
     nucleon remnant, also applies threshold factor for charm and 
     heavier quarks.  
\item[LREMH    \hfill (S)]
     gives energy-momentum fraction for target remnant split,
     cf.\ LST(14); also selects remnant flavours (not for PS case).
\item[LPRIKT   \hfill (S)]
     generates magnitude and azimuthal angle for Gaussian
     primordial $k_{\perp}$ of parton in nucleon, see PARL(3).
\item[LFRAME   \hfill (S)]
     transforms event to different frames, see above.
\item[LWBB     \hfill (S)]
     selects energy from a neutrino wide band beam, see above.
\item[LWEITS   \hfill (S)]
     integrates first order QCD matrix elements to set up grid in $x,W$ or 
     $x,y$ of weights for gluon radiation and boson-gluon fusion and finds
     maxima used for QCD simulation, see LINIT and common LGRID.
\item[LPRWTS   \hfill (S)]
     prints QCD weights, see above.
\item[LSIGMX   \hfill (S)]
     called by {\sc Minuit} routines to
     calculate differential electroweak cross section divided by
     the optimization function to obtain the maximum for the weighting
     procedure.
\item[LXSECT   \hfill (S)]
     called from LINIT to integrate cross section using {\sc Gadap}, 
     {\sc Riwiad} or {\sc Divonne}, see LINIT and LST(10).
\item[RIWIBD   \hfill (S)]
     substitutes block data for {\sc Riwiad} \cite{RIWIAD}, print flag changed.
\item[DVNOPT   \hfill (S)]
     substitutes block data for {\sc Divonne} \cite{DIVONNE}, 
     print flag changed.
\item[DFUN     \hfill (F)]
     integrand for {\sc Divonne} integration, calls RIWFUN.
\item[RIWFUN   \hfill (F)]
     integrand for {\sc Riwiad} integration, calls DCROSS.
\item[DCROSS   \hfill (F)]
     differential electroweak cross section used as integrand for
     numerical integration, see LXSECT.
     Integration variables given by LST(1), function is zero
     outside the cuts in array CUT.
\item[DLOWER, DUPPER   \hfill (F)]
     lower and upper limit on second kinematic variable
     ($y$,$Q^2$ or $W^2$)
     given a value of the first variable ($x$) and the cuts. Used
     together with DCROSS for cross section integration.
\item[FLTABL   \hfill (S)]
     called from LINIT to tabulate integrals in $x,Q^2$-grid for the
     longitudinal structure function, see LST(11).
\item[FLIPOL   \hfill (S)]
     called from LEPTOX to obtain longitudinal structure function from
     interpolation on grid from FLTABL, see LST(11).
\item[FLINTG   \hfill (S)]
     called from LEPTOX to obtain longitudinal structure function by
     integration event-by-event, see LST(11).
\item[FLQINT   \hfill (F)]
     integrand for quark contribution to QCD longitudinal structure
     function.
\item[FLGINT   \hfill (F)]
     integrand for gluon contribution to QCD longitudinal structure
     function.
\item[FLTINT   \hfill (F)]
     integrand for target mass correction, see LST(11).
\item[LSCI \hfill (S)] 
     generates soft colour interactions (SCI), see section 2.8. 
\item[LECSWI,LEASWI  \hfill (S)]
     switches colour and anticolour pointers for SCI. 
\item[LSMALL   \hfill (S)]
     creates hadrons from small mass colour singlet systems including diquarks.
\item[LSHOWR \hfill (S)]
     administer parton cascade evolution added to a $q$-event.
\item[LMEPS \hfill (S)]
     administer parton cascade evolution added to
     a $qg$- or $q\bar{q}$-event from first order QCD matrix elements.
\item[LSCALE \hfill (S)]
     gives scale for maximum virtuality in parton showers, see LST(9).
\item[LYSSPA \hfill (S)]
     simulate initial state parton cascade evolution,
     modification \cite{PS} of routine in \cite{PYTHIA48}.
\item[LYREMN, LYSPLI \hfill (S)]
     treatment of target remnant and primordial $k_\bot$ when using
     parton showers, modifications \cite{PS} of routines in
     \cite{PYTHIA48}.
\item[LMCMND,LMINTO,LMIDAT,LMINEW,LMPRIN,LMPINT,LMRAZZ,LMSIMP]
     modi\-fied {\sc Minuit} routines \cite{MINUIT} to find maximum of
     cross section.
\item[GADAP, GADAP2, GADAPF]
     one- and two-dimensional adaptive Gaussian integration routines
     \cite{GADAP}.
\item[LNSTRF   \hfill (S)]
     gives parton distributions in a nucleus, see above.
\item[LYSTFU \hfill (S)]
     interface to subroutine PYSTFU in {\sc Pythia} 5.7 \cite{JETSET}
     to give parton distribution functions, see above.
\end{defl}

\subsection{Common blocks}
Most of the communication between the user and the program is via
the switches and parameters in the common blocks.
The user need mainly be concerned with common LEPTOU since all
others are essentially for internal use.
All variables are given sensible default values in block data LEPTOD,
as shown by (D=...) below. These values may be changed by the
user to modify the behaviour of the program. Note, however, that
this should usually be done before calling LINIT and that some
of the parameters are interrelated.
Variables whose name start with D are in double precision.
The generated event is stored in common LUJETS, described in
\cite{JETSET}.

\noindent
COMMON /{\bf LEPTOU}/ CUT(14),LST(40),PARL(30),X,Y,W2,Q2,U
\begin{defl}{123456789012}
\item[{\it Purpose:}]
     contains input switches (LST(1)--LST(20),LST(34),LST(35)) 
     and input parameters
     (PARL(1)--PARL(20)) to specify physics, kinematic cuts and
     numerical procedures, as well as output flags (LST(21)--LST(40))
     and output variables (PARL(21)--PARL(30)).
     Overwriting default values should be made before calling LINIT.
\item[{\it Parameters:}]
\item[CUT(1), CUT(2)~~~:]
     (D$=10^{-4},1.$) lower and upper limit of Bjorken-$x$ variable.
\item[CUT(3), CUT(4)~~~:]
     (D$=0.,1.$) lower and upper limit of $y$ variable.
\item[CUT(5), CUT(6)~~~:]
     (D$=4.,10^8$) lower and upper limit of $Q^2$ (GeV$^2$).
\item[CUT(7), CUT(8)~~~:]
     (D$=5.,10^8$) lower and upper limit of $W^2$ (GeV$^2$).
\item[CUT(9), CUT(10)~~:]
     (D$=1.,10^8$) lower and upper limit of variable $\nu$ (GeV).
\item[CUT(11), CUT(12)~:]
     (D=$1.,10^8$) lower and upper limit of scattered lepton
     energy (GeV, in frame defined by LINIT call).
\item[CUT(13), CUT(14)~:]
     (D=$0.,3.1416$) lower and upper limit of lepton scattering angle
     (in radians), with respect to incoming lepton in frame
     defined by LINIT call.
\item[{\hfill {\it Remarks:}}]
     These cuts are applied already when choosing kinematic variables,
     before evaluating cross section formulae and
     structure functions, and will therefore be more efficient than
     applying cuts later on in the users program. The cross section
     estimates take these cuts into account.
     CUT(11)--CUT(14) are less efficient, since they refer to a special
     frame and the user should, if possible, translate them into cuts in
     $x,y,Q^2$.
\\
\item[LST(1)~~:]
     (D=0) choice of the two independent variables to be used for
     simulation and numerical integration of cross section.
\item[{\hfill =0:}]
     program makes a `best' choice for simulation efficiency depending on which
     process is to be generated (1 if $\gamma$-exchange included,
     else 2) cf.\ LST(31).
\item[{\hfill =1:}]  $x$ and $Q^2$
\item[{\hfill =2:}]  $x$ and $y$
\item[{\hfill =3:}]  $x$ and $W^2$
\item[LST(2)~~:]
     (D=1) choice of simulation and applying cuts in array CUT.
     (Used internally also with negative values.)
\item[{\hfill =1:}]
     kinematic   variables, LST(1), chosen from
     differential cross section and cuts applied.
\item[{\hfill =2:}]
     variables ($x,y$) supplied by user via LEPTOU, cuts applied.
\item[{\hfill =3:}]
     variables ($x,y$) supplied by user via LEPTOU, cuts not applied.
\item[LST(3)~~:] (D=5) regulates output and error handling.
\item[{\hfill =0:}] no output, execution not stopped on error.
\item[{\hfill =1:}] only warnings printed, execution not stopped on error.
\item[{\hfill =2:}] as 1, but execution stopped on error.
\item[{\hfill =3:}] as 2, and output at first initialization, no {\sc Minuit} 
     output.
\item[{\hfill =4:}] as 3, but output at all initialization calls.
\item[{\hfill =5:}] full output, i.e.\ as 4 and {\sc Minuit} output.
\item[LST(4)~~:]
     (D=1) regulates information in the event record. To be given as
     $I_{lepton} + 10\times I_{shower}$, where $I_{lepton}=0/1$
     inactives/actives the scattered lepton (K(i,1)=21/1)
     and $I_{shower}=0/1$ excludes/includes intermediate
     partons in the parton showers.
\item[LST(5)~~:] (D=3) choice of frame for the event.
\item[{\hfill =1:}] hadronic CM frame, $z$-axis along exchanged boson.
\item[{\hfill =2:}] lepton-nucleon CM frame, $z$-axis along lepton.
\item[{\hfill =3:}] lab system as defined by last user call to LINIT.
\item[{\hfill =4:}] as 3, but $z$-axis along exchanged boson.
\item[LST(6)~~:]
     (D=1) regulates the azimuthal angle,
      $\phi$, of the lepton scattering plane.
\item[{\hfill =0:}] no $\phi$-rotation, scattering plane is $x-z$ plane.
\item[{\hfill =1:}] random $\phi$-rotation,
     performed in lepton-nucleon cms for LST(5)$\geq 2$.
\item[LST(7)~~:] (D=1) regulates completeness of Monte Carlo simulation
     (to speed up program when only partial information is needed).
\item[{\hfill $=-1$:}] only kinematic variables generated.
\item[{\hfill $=0$:}]  kinematic variables and parton level event
     generated, optionally including QCD effects (cf.\ LST(8)).
     Hadronization can be made later by calling LUEXEC.
\item[{\hfill $=1$:}]
     full event generated, i.e.\ as 0 plus hadronization and decays.
\item[LST(8)~~:] (D=12) simulation of QCD effects in hadronic final
     state.
\item[{\hfill =0:}] QCD switched off.
\item[{\hfill =1:}] first order QCD matrix elements (ME)
     for gluon radiation and boson-gluon fusion.
\item[{\hfill =2:}]
     QCD parton cascade evolution from initial and final quark.
\item[{\hfill =3:}]
     QCD parton cascade evolution from initial quark only.
\item[{\hfill =4:}]
     QCD parton cascade evolution from final quark only.
\item[{\hfill =5:}]
     QCD switched off, but target remnant treatment as in 
     cascade case.
\item[{\hfill =9:}] set by {\sc Ariadne} \cite{ARIADNE} when simulating
     parton emission in the colour dipole model \cite{CDM}.
\item[{\hfill =12--15:}]
     as 2--5, but PS added on event from first order matrix elements.
\item[{\hfill {\it Remarks:}}]
     See PARL(8),PARL(9) for cut-off on ME. Without PS one may decrease
     the cutoff to maximize QCD emission.
\item[LST(9)~~:] (D=5) scale in parton showers, i.e. maximum virtuality
      (mass-squared) of parton initiating the shower. Only used when
     QCD ME are not included, i.e.~LST(8)=2,3,4. See further section 2.6.
\item[{\hfill =1:}] $Q^2$
\item[{\hfill =2:}] $W^2$
\item[{\hfill =3:}] $W*Q$
\item[{\hfill =4:}] $Q^2(1-x)$, which is
     $\sim \langle p_{\bot}^2\rangle$ from ME at large $x$
\item[{\hfill =5:}] $Q^2(1-x)\, max(1,ln\frac{1}{x})$, which is
     a combination of $\langle p_{\bot}^2\rangle$-dependence from ME at
     large and small $x$
\item[{\hfill =6:}] $x_0W^2$, to represent the mass-squared of the
     hadronic system whithout the (non-perturbative) spectator.
     A valence-like parton distribution $f(x_0,Q^2_0)\sim (1-x_0)^a$
     ($a=4$) is used to chose $x_0$ representing an original parton.
\item[{\hfill =9:}] $W^{4/3}$, i.e.~similar as in the colour dipole
     cascade model \cite{ARIADNE}.
\item[LST(10)~:]  (D=1) numerical integration of cross section in LINIT.
\item[{\hfill =0:}] not performed.
\item[{\hfill =1:}]
     performed using {\sc Gadap} \cite{GADAP}, i.e.\ adaptive Gaussian method.
\item[{\hfill =2:}] performed using {\sc Riwiad} \cite{RIWIAD}, 
     i.e. adaptive Monte Carlo method.
\item[{\hfill =3:}]
     performed using {\sc Divonne} \cite{DIVONNE}, automatic invocation.
\item[{\hfill =4:}]
     performed using {\sc Divonne} \cite{DIVONNE}, detailed  invocation.
\item[{\hfill {\it Remarks:}}]
     Integration variables are defined by LST(1), integration region by
     the cuts in array CUT and required accuracy by PARL(15).
     Result stored in PARL(23). Under normal conditions the simpler
     {\sc Gadap} routine is accurate enough.
\item[LST(11)~:]
     (D=0) inclusion of longitudinal structure function,
     target mass and higher twist corrections, see section 2.3, 
     for $\gamma$ and $\gamma /Z$ exchange. To be set as
     LQCD+10*LTM+100*LHT where LQCD, LTM and LHT are 0 or 1 to
     exclude or include contributions from QCD, Target Mass and Higher Twist,
     respectively. The QCD and target mass parts involve integrals
     which are evaluated at initialization and stored on an $x,Q^2$
     grid used for interpolation when simulating events. In this mode,
     the cross section estimates PARL(23) and PARL(24) include $F_L$.
     By setting
     LQCD or LTM equal 2, the corresponding contribution are
     evaluated by integration for each event at the proper $x,Q^2$
     point, which gives a more accurate result. In this case, the 
     result is included in PARL(24) but not in PARL(23) since nested 
     integrations is too time consuming.
\item[LST(12)~:]
     (D=4) maximum flavour used in sea structure function parametrizations.
\item[LST(13)~:]  (D=5) heaviest quark flavour allowed in boson-gluon
                  fusion. A threshold factor is applied to compensate
                  for neglected quark masses in the matrix elements.
\item[LST(14)~:]  (D=4) treatment of target remnant after removing
     interacting parton, see section 2.7.
\item[{\hfill =0:}]
     remnant approximated by anti-parton of removed parton,
     i.e.\ by $q,\bar{q},g$ for removed $\bar{q},q,g$.
     No baryon is produced.
\item[{\hfill =1:}]
     for removed valence quark the remnant is a diquark hadronizing
     into a baryon with the Lund model.
     For a removed gluon, sea quark ($q_s$), sea antiquark the remnant
     is,
     respectively, a $q_vq_vq_v$, $q_vq_vq_v\bar{q}_s$, $q_vq_vq_vq_s$
     which is split into $q_vq_v+q_v$, $q_vq_v+M(q_v\bar{q}_s)$,
     $q_v+B(q_vq_vq_s)$. The (lighter) part of the remnant containing
     one random flavour valence quark $q_v$, takes the energy-momentum
     fraction $z$ given by $P(z)=2(1-z)$, i.e. $\langle z \rangle =1/3$.
\item[{\hfill =2:}]
     as 1, but with $P(z)=(a+1)(1-z)^a$ with $a$ chosen such that
     $\langle z \rangle =1/(a+2)=m/(m+M)$ where $m$ ($M$) is the mass
     of the light (heavy) remnant subsystem.
\item[{\hfill =3:}]
     as 2, but using the `Peterson' function
     $P(z)=N/(z(1-1/z-c/(1-z))^2)$ with $c=(m/M)^2$.
\item[{\hfill =4:}]
     using LUZDIS, i.e. the fragmentation function chosen in {\sc Jetset}. 
\item[LST(15)~:]  (D=9) 
     choice of parton distribution functions, $xq(x,Q^2)$ and $xg(x,Q^2)$,
     for the target hadron, see section 2.4 and LST(16).
\item[{\hfill =0:}]  parton density choice and parameters are controlled
     directly through parameters MSTP(51), 
     MSTP(52), MSTP(57), MSTP(58) and PARP(51) in common PYPARS 
     in {\sc Pythia} 5.7 \cite{JETSET}.
\item[{\hfill =1:}]  Eichten-Hinchliffe-Lane-Quigg set 1 \cite{EHLQ}.
\item[{\hfill =2:}]  Eichten-Hinchliffe-Lane-Quigg set 2 \cite{EHLQ}.
\item[{\hfill =3:}]  Duke-Owens set 1 \cite{DO}.
\item[{\hfill =4:}]  Duke-Owens set 2 \cite{DO}.
\item[{\hfill =5:}]  CTEQ2M (best $\overline{\mbox{MS}}$ fit) \cite{CTEQ}.
\item[{\hfill =6:}]  CTEQ2MS (singular at small-$x$) \cite{CTEQ}.
\item[{\hfill =7:}]  CTEQ2MF (flat at small-$x$) \cite{CTEQ}.
\item[{\hfill =8:}]  CTEQ2ML (large $\Lambda$)) \cite{CTEQ}.
\item[{\hfill =9:}]  CTEQ2L (best leading order fit) \cite{CTEQ}.
\item[{\hfill =10:}]  CTEQ2D (best DIS fit) \cite{CTEQ}.
\item[{\hfill =-4:}]  Intrinsic charm quarks only, see section 2.4;
     to be used with PARL(3)$\simeq m_c\simeq 1~GeV$ and
     overall normalisation PARL(12). 
\item[{\hfill =-5:}]  Intrinsic bottom quarks only, approximated
     by taking the same distribution as for intrinsic charm, but
     renormalized by $m_c^2/m_b^2\simeq 0.1$.
\item[{\hfill {\it Remarks:}}]
     Except for intrinsic charm and beauty, 
     the parton densities are obtained from subroutine PYSTFU in 
     {\sc Pythia} 5.7 \cite{JETSET}; with parameters set to MSTP(51)=LST(15),
     MSTP(52)=LST(16) and MSTP(58)=LST(12), unless LST(16)=0 when the 
     user directly controls the PYSTFU parameters. 
     To evaluate structure functions separately, see
     subroutines LYSTFU above.
\item[LST(16)~:]  (D=1) choice of proton parton-distribution-function library,
     i.e. MSTP(52) in {\sc Pythia} 5.7 \cite{JETSET}.
\item[{\hfill =1:}]  internal {\sc Pythia} according to LST(15).
\item[{\hfill =2:}]  PDFLIB (version 4 or later) \cite{PDFLIB}, with 
     with NGROUP and NSET to be given as MSTP(51)=1000*NGROUP+NSET.
\item[LST(17)~:]  (D=0) regulates varying energies of initial particles
     from event to event.
\item[{\hfill =0:}]  fixed energies as specified in LINIT.
\item[{\hfill =1:}]
     energies allowed to vary; momenta should be given in P(i,j) with
     i=1,2 for lepton, nucleon and j=1...5 for $p_x,p_y,p_z,E,m$.
     See also LWBB.
     Note: this option is not fully tested, use ME only with LST(19)=-1.
\item[LST(18)~:]  (D=2) running of electromagnetic coupling $\alpha$
     and choice of $W$ and $Z$ masses.
\item[{\hfill =0:}]
     $\alpha$ fixed to value at $Q^2=0$ given in PARL(16), $Z$ and $W$
     masses given independently by PMAS(23), PMAS(24) in /LUDAT2/.
\item[{\hfill =1:}]
     as 0, but $W,Z$ masses calculated from standard model using
     $\sin{\theta_W}, \alpha , G_F$ and radiative corrections;
     i.e.\ PARL(5), PARL(16), PARL(17), PARL(18)
\item[{\hfill =2:}]
     as 1, but $\alpha$ varying with $Q^2$ using ULALEM
     in \cite{JETSET}, see MSTU(101).
\item[LST(19)~:]  (D=-10) regulates use of grid to store probabilities
     for $qg$- and $q\bar{q}$-events.
\item[{\hfill =-10:}] the probabilites are obtained from an automatic grid, but
close to the saturation limit they are recalculated to avoid interpolation
errors.
\item[{\hfill =-1:}] no grid, necessary integrals calculated for each
     event, more time consuming but higher precision.
\item[{\hfill =0:}]
     user defined grid to be read in free format, see comments in
     subroutine LWEITS.
\item[{\hfill =1:}]
     grid suitable for lepton beam energy $<300$ GeV on fixed target.
\item[{\hfill =2:}]
     grid suitable for lepton beam energy $<1000$ GeV on fixed target.
\item[{\hfill =3:}]  grid suitable for $ep$ collisions in HERA.
\item[{\hfill =4:}]  grid suitable for $ep$ collisions in LEP+LHC.
\item[{\hfill =10:}] grid chosen automatically after the kinematic
     region specified by user.
\item[LST(20)~:] (D=5) scheme for cut-offs against divergences in
     the QCD matrix elements \cite{mixed}, {\it cf.} section 2.5 and 
     PARL(8),PARL(9).
     ($m_{ij}$ is the invariant mass of any parton pair).
\item[{\hfill =1:}] $W^2$ (or JADE) scheme, $m_{ij}^2>y_{cut}W^2$
\item[{\hfill =2:}] $Q^2$ scheme, $m_{ij}^2>y_{cut}Q^2$
\item[{\hfill =3:}] mixed scheme, i.e.
     $m_{ij}^2>c_1Q^2$ for partons $i,j$ that are not spectators and 
     $2p_i\cdot P>c_2Q^2/x=c_2(W^2+Q^2)$ else.
     Virtuality scale for matched parton showers in $q$-event is
     $c_1Q^2$.
\item[{\hfill =4:}] as 3, but with virtuality $c_2Q^2/x$ for initial
     state parton showers.
\item[{\hfill =5:}] $z-\hat{s}$ scheme, 
     $\hat{s}>\hat{s}_{min}$ and $z_q<z_{q,min}<1-z_q$. To regulate the 
     divergence, $z_q$ is changed, see PARL(27).
\item[{\hfill =6:}] as 5, but $\hat{s}_{min}$ is changed to regulate 
     the divergence, see PARL(27).
\item[LST(21)~:]
     error flag, =0 for properly generated event. Nonzero value
     indicates incorrect event that should be rejected;
     may occur in case of user supplied kinematic variables or variable
     beam energies, cf.\ LST(2) and LST(17).
\item[LST(22)~:]  specifies chosen target nucleon in current event.
\item[{\hfill =1:}]  proton.
\item[{\hfill =2:}]  neutron.
\item[LST(23)~:]  specifies process simulated.
\item[{\hfill =1:}]  electromagnetic (EM), i.e.\ $\gamma$ exchange.
\item[{\hfill =2:}]  weak charged current (CC), i.e. $W^{\pm}$ exchange.
\item[{\hfill =3:}]  weak neutral current, i.e.\ $Z^0$ exchange.
\item[{\hfill =4:}]  neutral current (NC), i.e.\ $\gamma /Z^0$ exchange.
\item[LST(24)~:]  specifies first order QCD process in current event.
\item[{\hfill =1:}] $q$-event, i.e.\ no first order QCD.
\item[{\hfill =2:}] $qg$-event, i.e. gluon radiation in first order QCD.
\item[{\hfill =3:}] $q\bar{q}$-event, i.e.\ boson-gluon fusion in first
     order QCD.
\item[LST(25)~:] specifies flavour of struck quark in current event:
     1=$d$, 2=$u$, 3=$s$, 4=$c$, 5=$b$, --1=$\bar{d}$, --2=$\bar{u}$,
     --3=$\bar{s}$, --4=$\bar{c}$, --5=$\bar{b}$.
\item[LST(26)~:] entry line in event record of outgoing struck quark.
     In parton shower case, quark at boson vertex before final state shower.
\item[LST(27)~:]  signals split of non-trivial nucleon remnant,
     cf.\ LST(14).
\item[{\hfill =0:}]  no split, simple diquark or LST(14)=0.
\item[{\hfill =1:}]  split into parton and particle, $qq+M$ or $q+B$,
     occurs when sea (anti)quark removed through the interaction.
\item[{\hfill =2:}]  split into quark and diquark, $q+qq$,
     occurs when a gluon is removed.
\item[LST(28)~:]  specifies the frame in which the current event is
     given with code as for LST(5), cf.\ LFRAME.
\item[LST(29)~:] specifies azimuthal angle rotation with code as for
     LST(6), cf.\ LFRAME.
\item[LST(30)~:]  specifies chosen helicity of beam lepton in current
     event.
\item[{\hfill $=-1$:}] left-handed.
\item[{\hfill $=+1$:}] right-handed.
\item[LST(31)~:] specifies chosen variables for the simulation/integration.
\item[{\hfill =1:}]  $x$ and $Q^2$.
\item[{\hfill =2:}]  $x$ and $y$.
\item[{\hfill =3:}]  $x$ and $W^2$.
\item[LST(32)~:]  Internal flag, 0 or 1 for simulation and
     integration, respectively.
\item[LST(33)~:]  Reserved for internal test.
\item[LST(34)~:]  (D=1) switch for soft colour interactions, 
                  0=off, 1=on; see section 2.8. 
\item[LST(35)~:]  (D=1) switch for new sea quark treatment, 0=off, 1=on, 
                  see section 2.7. 
                  (Inactive for scattering on intrinsic charm.)
\item[LST(36)--LST(40)~:]  unused at present.
\\
\item[PARL(1)~~:] (D=1.) number of nucleons in target nucleus, i.e. $A$.
\item[PARL(2)~~:] (D=1.) number of protons in target nucleus, i.e.\ $Z$.
\item[PARL(3)~~:]  (D=0.44 GeV) width of Gaussian distribution for the
     primordial transverse momentum $k_\bot$ of partons in the nucleon.
\item[PARL(4)~~:]  (D=0.75) probability that a $ud$-diquark in the
     target remnant has spin and isospin equal zero, i.e.\ I=S=0.
\item[PARL(5)~~:]  (D=0.2319) $\sin^2{\theta_W}$ (Weinberg angle)
     \cite{PDG}.
\item[PARL(6)~~:]  (D=0.) polarization of lepton beam; should be set to
     $1-2P_L=2P_R-1$ where $P_{L,R}$ is the probability of left(right)-handed
     helicity, i.e. $-1(+1)$ corresponds to a fully left(right)-handed polarized
     beam.
\item[PARL(7)~~:]  (D=0.5) probability for soft colour interactions 
     between parton pairs, see section 2.8. 
\item[PARL(8),PARL(9)~:]
     (D=0.04,4) cut-offs against divergences in the QCD matrix elements, cf.
     section 2.5 and LST(20). Suitable values are given in parenthesis. \\
     LST(20)=1,2: PARL(8)$=y_{cut}$ and PARL(9)=$\min m_{ij}$ in GeV. (0.005,2)\\
     LST(20)=3,4: PARL(8)$=c_2$ and PARL(9)$=c_1$. (0.05,2)\\
     LST(20)=5,6: PARL(8)$=z_{q,min}$ and PARL(9)$=\hat{s}_{min}$ in GeV$^2$.
     (0.04,4) 
\item[{\hfill {\it Remarks:}}]
     These are starting values when integrating first order QCD matrix
     elements, but the effective cut used, PARL(27), is automatically increased
     in order that the QCD probabilities do not exceed unity.
\item[PARL(10)~:] Not used at present.
\item[PARL(11)~:]
     (D=0.01) required relative accuracy for one-dimensional
     integration, used for first order QCD matrix element weights
     and longitudinal structure function integrals.
\item[PARL(12)~:] (D=0.01) probability $\beta^2$ for an intrinsic charm
     quark-antiquark pair in the proton (cf.\ section 2.4). 
\item[PARL(13)~:]
     (D=0.1) internal parameters used for adjustment of $y_{cut}$
     for integration of QCD matrix elements.
\item[PARL(14)~:]
     (D=0.35 GeV) width of Gaussian distribution in transverse momentum
     when a non-trivial target remnant is split into two particles, 
     cf.\ LST(27).  
\item[PARL(15)~:]
     (D=0.01) required relative accuracy for two-dimensional integration
     to get total cross section, cf.\ LST(10).
\item[PARL(16)~:]
     (D$=7.29735\times 10^{-3}$) finestructure constant $\alpha$
     \cite{PDG}, cf.\ LST(18).
\item[PARL(17)~:]
     (D$=1.16639\times 10^{-5}$ GeV$^{-2}$)
     weak Fermi constant $G_F$ \cite{PDG}.
\item[PARL(18)~:]  (D=0.044) $\Delta r$ from radiative corrections \cite{PDG}.
\item[PARL(19)~:]  (D=0.03) scale $\kappa^2$ in GeV$^2$
     for higher twist correction \cite{HTWIST}, see eq. (\ref{FLHT}).
\item[PARL(20)~:] (D=0.1 GeV) in case of a more complicated nucleon 
     remnant than a simple diquark, the constituent masses of quarks 
     and diquarks in the remnant are decreased with PARL(20) and 
     2*PARL(20) respectively.
\item[PARL(21)~:]
     $2P\cdot k$, where $P$ is proton and $k$ lepton four-vectors,
     equals invariant mass squared, $s$, when masses are neglected.
\item[PARL(22)~:]
     $2P\cdot q$, where $P$ is proton and $q$ is boson four-vectors.
\item[PARL(23)~:]
     cross section in $pb$, corresponding to the kinematic region
     allowed by the cuts in array CUT, obtained by numerical
     integration in LINIT call, cf.\ LST(10).
     If LQCD=2 or LTM=2, cf.\ LST(11), the corresponding contribution
     to $F_L$ is not included.
\item[PARL(24)~:]
     Monte Carlo estimate of the cross section in $pb$ associated with
     the generated event sample, taking CUT into account. Set to
     zero by LINIT call and updated with each event generated, hence
     accuracy improves with statistics as $1/\sqrt{N}$ and only final
     value should be used.
\item[PARL(25)~:]  value of $\alpha_s$ in current event.
\item[PARL(26)~:]  value of $\Lambda$ obtained from structure functions, which  
     is used in $\alpha_s$ and in the initial state parton shower.  
\item[PARL(27)~:] depending on LST(20), 
     present value of the $y_{cut}$-, $z_q$- or $\hat{s}$-cut for first order 
     QCD. These are given by PARL(8) and PARL(9), but modified internally to 
     prevent QCD weights to exceed unity.
\item[PARL(28),PARL(29),PARL(30)~:] values of $x_p,z_q$ and $\phi$ in first 
     order massless QCD matrix elements, section 2.5, for current event 
     if it is a $qg$- or $q\bar{q}$-event, see LST(24). For a $q-$event they are
     set to $1.0$, $1.0$ and $0.0$.
\\
\item[X~~:]  Bjorken-$x$, i.e.\ $x=Q^2/2P\cdot q$.
\item[Y~~:]  standard $y$ variable, i.e.\ $y=P\cdot q / P\cdot p_e$.
\item[W2~:]  mass-square of hadronic system, i.e.\ $W^2=(P+q)^2$.
\item[Q2~:]  momentum transfer squared,
     i.e.\ $Q^2=-q^2=-(p_e-p_{\ell})^2$.
\item[U~~:]  energy transfer variable  $\nu = P\cdot q/m_p$.
\item[{\hfill {\it Remark:}}] for details see section 2.1.
\end{defl}

\noindent
COMMON /{\bf LFLMIX}/ CABIBO(4,4)
\begin{defl}{123456789012}
\item[{\it Purpose:}]
     Contains the Cabbibo-Kobayashi-Maskawa matrix elements squared
     for flavour mixing. First index corresponds to
     the up-quark flavours $u,c,t,h$ and second index to the down-quark
     flavours $d,s,b,l$. The default values are \cite{PDG}
     CABIBO/.95,.05,2*0.,.05,.948,.002,2*0.,.002,.998,4*0.,1./
\end{defl}

\noindent
COMMON /{\bf LOPTIM}/ OPTX(4),OPTY(4),OPTQ2(4),OPTW2(4),COMFAC
\begin{defl}{123456789012}
\item[{\it Purpose:}]
     parameters to optimize simulation efficiency, see section 2.2.
     Default values are good for normal usage, but improvements may be
     possible under, e.g., unusual kinematic conditions.
     For changes, see their detailed meaning given by
     comments in the code of subroutine LEPTOX.
\end{defl}

\noindent
COMMON /{\bf LBOOST}/ DBETA(2,3),STHETA(2),SPHI(2),PB(5),PHIR
\begin{defl}{123456789012}
\item[{\it Purpose:}] rotation and boost parameters for transforming event
     between frames. DBETA(i,j), STHETA(i) and SPHI(i) are the boosts,
     polar and azimuthal angles for transforming from lepton-nucleon
     cms to
     lab (i=1) and from hadronic cms to lepton-nucleon cms (i=2).
     For these cases, the order of the transformations should be
     first rotation in $\theta$, then in $\phi$ followed by the boost
     where j=1,2,3 corresponds to $x,y,z$ components. Should only be
     used via subroutine LFRAME to set flags properly, cf.\ LST(5) and
     LST(6).
\end{defl}

\noindent
COMMON /{\bf LGRID}/ NXX,NWW,XX(31),WW(21),PQG(31,21,3),PQQB(31,21,2),\\
\& QGMAX(31,21,3),QQBMAX(31,21,2),YCUT(31,21),XTOT(31,21),NP
\begin{defl}{123456789012}
\item[{\it Purpose:}]
     information for simulating first order QCD processes.
\item[{\it Remarks:}] Probabilities PQG and PQQB
     for gluon radiation and boson-gluon fusion, i.e.\ $qg$- and
     $q\bar{q}$-events, are stored on a grid in $x,W$ (or $x,y$ for
     LST(19)=10) with NXX, NWW points defined by XX and WW. Indices are
     for $x$, $W$ ($y$) and
     helicity contributions. The cut on parton pair masses $y_{ij}$ is
     also stored and maxima for the Monte Carlo rejection technique.
     The grid content is set up by LWEITS and can be printed by LPRWTS.
     See LST(19) and section 2.5 for physics and methods.
\end{defl}

\noindent
COMMON /{\bf FLINFO}/ RFLQ,RFLG,RFLM,RFLT
\begin{defl}{123456789012}
\item[{\it Purpose:}]
     gives the relative contributions from the different parts
     (quark, gluon, mass, higher twist)
     of the longitudinal structure funtion to the differential cross
     section at the $x,Q^2$ point of the current event,
     cf.\ section 2.3.
\end{defl}

\noindent
COMMON /{\bf LYPARA}/ IPY(80),PYPAR(80),PYVAR(80)
\begin{defl}{123456789012}
\item[{\it Purpose:}]
     parameters for parton cascade routines. This is common PYPARA
     from {\sc Pythia} 4.8 and described in detail in \cite{PYTHIA48}.
     Only those parameters used with {\sc Lepto} in parton cascade mode are
     commented here (default values as in {\sc Pythia} 4.8 if not given).
\item[{\it Parameters:}]
\item[IPY(8)~~:]  set to LST(12) in LINIT.
\item[IPY(13)~:]  set to zero in LINIT if LST(8)=3,5,13,15.
\item[IPY(14)~:]  set to zero in LINIT if LST(8)=4,5,14,15.
\item[IPY(11),IPY(15),IPY(34),IPY(40)--IPY(42),IPY(47),IPY(48)~:]
     are used.
\item[PYPAR(8),PYPAR(11)--PYPAR(16)~:]  are used.
\item[PYPAR(21)~:] (D=PARL(26)) $\Lambda_{QCD}$ in initial state parton shower.
\item[PYPAR(22)~:] (D=1. GeV$^2$) cutoff for initial state parton shower.
\item[PYPAR(23),PYPAR(24)~:]  are used.
\item[PYPAR(25),PYPAR(26)~:] (D=2*1.) multiplies the chosen scale,
     cf.\ LST(9), to give maximum virtuality for final and initial state
     showers, respectively.
\item[PYPAR(27)~:] (D=1.) multiplies virtuality $Q^2$ for $\alpha_s$
     and structure functions in initial state showers.
\item[PYVAR(1)--PYVAR(5)~:] are used.
\end{defl}

\noindent
COMMON /{\bf LUJETS}/ N,K(4000,5),P(4000,5),V(4000,5)\\
COMMON /{\bf LUDAT1}/ MSTU(200),PARU(200),MSTJ(200),PARJ(200)
\begin{defl}{123456789012}
\item[{\it Purpose:}]
     LUJETS contains the record of the generated event and is
     essential for using the results. LUDAT1 contains switches and
     parameters that are, e.g., essential to control final state parton
     showers, $\alpha_s$ evaluation and hadronization.
     These common blocks are described in the {\sc Jetset} manual 
     \cite{JETSET}.
\end{defl}

\noindent
Short description of all common blocks:
\noindent
\begin{defl}{123456789012}
\item[{\it Common}]  {\it Purpose}
\item[LEPTOU] main common block for user control of program, see above.
\item[LFLMIX] quark flavour mixing parameters (KM-matrix), see above.
\item[LOPTIM] parameters to optimize simulation efficiency, see above.
\item[LBOOST] boost and rotation parameters, see above.
\item[LGRID ] grid for first order QCD event simulation, see above.
\item[LINTER] internal parameters and variables; charges, couplings,
              cross section weights.
\item[LINTRL] internally used; basic system in different frames,
              effective limits on kinematic variables.
\item[LPFLAG] internal for output control.
\item[LINTEG] internal counters for integrand evaluations.
\item[FLGRID] grid with integrals for longitudinal structure function.
\item[FLINFO] relative size of longitudinal structure function contributions,
              see above.
\item[LYPARA] parameters for parton cascade evolution \cite{PYTHIA48}, 
              see above.
\item[LYPROC,LYINT1] internal for parton showers \cite{PYTHIA48}.
\item[LMINUI,LMINUC]
     input parameters and character names for {\sc Minuit} \cite{MINUIT}.
\item[LM....] internally used in adaptation of {\sc Minuit} routines.
\item[GADAP1] internal in {\sc Gadap} integration routines.
\item[PARAMS,ANSWER] input/output for {\sc Riwiad} integration \cite{RIWIAD}.
\item[STORE,STORE1,OPTION,RANDOM,INTERN]  internal for {\sc Riwiad} 
     \cite{RIWIAD}.
\item[BNDLMT,SAMPLE,PRINT] input for {\sc Divonne} integration \cite{DIVONNE}.
\item[ARDAT1] parameters in the {\sc Ariadne} MC for dipole radiation
              \cite{ARIADNE}.
\item[LUDAT1,LUDAT2] switches, parameters, particle data
     in {\sc Jetset} \cite{JETSET}.
\item[PYPARS] switches, parameters used for parton densities in 
     {\sc Pythia} 5.7 \cite{JETSET}.
\item[LUJETS] contains generated event, see \cite{JETSET}.
\end{defl}
\newpage
\section{Usage and availability}
{\sc Lepto} 6.5 should be loaded together with {\sc Jetset} 7.4 and 
{\sc Pythia} 5.7 \cite{JETSET} (the latter to access parton density 
parametrizations). 
The ordinary gamma function, GAMMA(X), is called and must be supplied
(usually available in FORTRAN77).
Access to the CERN library is not necessary, but gives access to
the {\sc Pdflib}, {\sc Riwiad} and {\sc Divonne} program packages as 
well as the TIMEX routine. 
The program is a slave system, which the user must call from
his own steering program.

Information about the program, its update history, source code, example jobs
etc.\ can be obtained on request from the authors or directly via the WWW on
the {\sc Lepto} home page, \texttt{ http://www3.tsl.uu.se/thep/lepto/}.

\vspace{5mm}
{\bf Acknowledgements:}
We are grateful to the many persons who have directly or indirectly
contributed to the developlement of this program. In particular,
thanks go to M. Bengtsson and T. Sj\"ostrand for collaboration on 
parton showers introduced in version 5.2.


\end{document}